\documentclass[a4paper,12pt]{article}
 \usepackage[latin1]{inputenc}
\usepackage[T1]{fontenc}
\usepackage{amsmath}
\usepackage{amsfonts}
\usepackage{slashed}
\usepackage{amssymb}
\usepackage{color}
\usepackage{graphicx}
\usepackage{color}
\usepackage{dsfont}
\usepackage{upgreek}
\usepackage{soul}
\usepackage{caption,subcaption}
\usepackage[normalem]{ulem}
 \newcommand{\be}{\begin{equation}}
  \newcommand{\bb}{\begin{equation}}
	 \newcommand{\ee}{\end{equation}}
	 \newcommand{\ba}{\begin{eqnarray}}

		 \newcommand{\ea}{\end{eqnarray}}
		 
		   \newcommand{\bea}{\begin{eqnarray}}
			 \newcommand{\eea}{\end{eqnarray}}
\newcommand{\eqb}{\begin{eqnarray}}
\newcommand{\eqf}{\end{eqnarray}}

\begin{document}
  \title{Visible and hidden sectors in a model with  Maxwell and Chern-Simons gauge dynamics}
\author{Edwin Ireson$^{\,a}$, Fidel A. Schaposnik$\,^b$ and Gianni Tallarita$\,^{c}$
~\\
~\\
{$^a\,$\it Department of Physics, Swansea University}\\
{\it Singleton Park, Swansea, SA2 8PP, UK}
\\
~\vspace{-3 mm}
\\
{$^b\,$\it Departamento de F\'\i sica, Universidad Nacional de La Plata/IFLP/CICBA} \\{\it CC 67, 1900 La Plata, Argentina}
\\
~\vspace{-3 mm}
\\
{$^c$ \it
Departamento de  Ciencias, Facultad de Artes Liberales}\\
{\it Universidad Adolfo Ib\'a\~nez
 Santiago 7941169, Chile.}
}
\maketitle

%===================================================================
\begin{abstract}
We study   a $U(1) \times U(1)$ gauge theory discussing its vortex solutions and supersymmetric extension.
In our set-upon the dynamics of one of two Abelian gauge fields is governed by a Maxwell term, the other by a Chern-Simons term. The two sectors interact
 via a BF gauge field mixing and a Higgs portal term that connects the two complex scalars. We also consider the supersymmetric version of this system which allows to find for the bosonic sector BPS equations in which an additional real scalar field enters into play.
We  study numerically the field equations finding  vortex solutions with both magnetic flux and electric charge.
\end{abstract}
%===================================================================

\section{Introduction}
Gauge fields coupled to charged scalars in  models with two weakly connected sectors have recently attracted much attention both in high energy  and condensed matter physics. Concerning the former, many extensions of the Standard Model (SM) propose the existence of additional product gauge groups in models where there are particles transforming under the new gauge fields and not under the usual local symmetries (See \cite{JR}-\cite{PA} and references therein). Also, different supersymmetry extensions of the SM include a hidden sector intended to break supersymmetry (SUSY) leading to acceptable superpartner masses. This is the case of the supersymmetric model proposed in \cite{dienes} with two $U(1)$ gauge symmetries in which the gauge field in the hidden sectors communicates SUSY breaking to the visible sector through a gauge kinetic term.

Two possible  gauge-invariant and renormalizable mixings between the fields of two Abelian Higgs model sectors can be considered, either coupling the gauge fields or the complex scalars.
In the former case working in $3+1$ space-time dimensions one has the so called gauge kinetic mixing with an interaction Lagrangian between the two sectors originally introduced in \cite{Okun}-\cite{Holdom}
\be
L_{mix} = \xi F_{\mu\nu}[A] F^{\mu\nu}[B]
\ee
where $F_{\mu\nu}$ is the field strength associated to each one of the two $U(1)$ gauge fields
$A_\mu, B_\mu$.

In the case  of $2+1$ space-time, relevant for describing systems in planar  physics, there is a second gauge mixing which takes the form of a BF term,
\be
L_{mix} = \xi'  \epsilon^{\mu\nu\alpha}B_\mu F_{\nu\alpha}[A]
\ee

Concerning the scalar-scalar mixing, one has the so-called  Higgs portal interaction  $L_{HP}$, originally introduced in refs. in \cite{SZ}-\cite{PW},
\be
L_{HP} = \zeta |\phi^2| |\eta|^2
\ee
with $\phi$ and $\eta$ the complex scalars. It is interesting to note that if one takes $\phi$ as the scalar doublet  of the Standard Model and $\eta$  as a Goldstone boson, a fractional increase in the number of relativistic species (dark radiation; e.g., neutrinos) can be produced \cite{Wei}.

Classical string-like solutions to the field equations of models with two Abelian Higgs sectors connected by a gauge mixing interaction have been found and their possible relevance in the context of dark matter has been discussed  \cite{HA}-\cite{AS}.

Supersymmetric extensions of gauge theories coupled to scalars having two sectors have been also considered in \cite{MS} in connection with lower-energy precision experiments. Now, supersymmetry necessarily require both types of mixing with couplings appropriately related, precisely in the way required for the  existence of self-dual first order (BPS) equations in the bosonic sector \cite{deVS}-\cite{ENS}.  In the case of models with visible and hidden sectors   this has been studied  for a $3+1$ dimensional $U(1)\times U(1)$ model with   gauge dynamics  governed by a Maxwell term in each sector  in ref.\cite{AINS} and  for a $2+1$ dimensional model with Chern-Simons gauge dynamics \cite{AIST}.

Regarding condensed matter applications,  $2 +1$ dimensional Abelian Higgs models with two sectors, in one of which the $U(1)$ symmetry remains unbroken,  interesting effects in connection with superconductivity were discussed  \cite{AS},\cite{Anber}.

It is the purpose of the present work to consider the case in which the gauge dynamics is a $2 +1$ dimensional model  in which gauge dynamics in each sector corresponds to different kinetic terms, namely  {a} Maxwell kinetic term in one sector and   {a} Chern-Simons  {term} in the other. The model could be of interest both in the analysis of a $3+1$ field theory at high temperature and also in the study of parity violation phenomena in planar physics.

The plan of the paper is the following.
We first consider   (in section 2) a  bosonic   $U(1) \times U(1) $ gauge theory with spontaneous symmetry breaking  in both sectors which are connected  {through} BF gauge mixing  and Higgs portal interactions. We then  proceed in section 3 to a SUSY extension of the model {,} finding  first order self-dual equations associated   {with} the bosonic sector. In section 4 we analyze the numerical solutions of both models, finally giving  in section 5 a summary of the results and conclusions.

%%%%%%%%%%%%%%%%%%%%%%%%%%%%%%%%%%%%%%%%%%%%%%%%%%%%%%%%%%%%%%%%%%%%%%%%%%%%%%%%%%%%%%%%%%%%%%%%%%%%%%%
%%%%%%%%%%%%%%%%%%%%%%%%%%%%%%%%%%%%%%%%%%%%%%%%%%%%%%%%%%%%%%%%%%%%%%%%%%%%%%%%%%%%%%%%%%%%%%%%%%%%%%%
%%%%%%%%%%%%%%%%%%%%%%%%%%%%%%%%%%%%%%%%%%%%%%%%%%%%%%%%%%%%%%%%%%%%%%%%%%%%%%%%%%%%%%%%%%%%%%%%%%%%%%%
%%%%%%%%%%%%%%%%%%%%%%%%%%%%%%%%%%%%%     SECTION 2       %%%%%%%%%%%%%%%%%%%%%%%%%%%%%%%%%%%%%%%%%%%%%
%%%%%%%%%%%%%%%%%%%%%%%%%%%%%%%%%%%%%%%%%%%%%%%%%%%%%%%%%%%%%%%%%%%%%%%%%%%%%%%%%%%%%%%%%%%%%%%%%%%%%%%
%%%%%%%%%%%%%%%%%%%%%%%%%%%%%%%%%%%%%%%%%%%%%%%%%%%%%%%%%%%%%%%%%%%%%%%%%%%%%%%%%%%%%%%%%%%%%%%%%%%%%%%
\section{The $U(1)\times U(1) $ model}
We shall consider a $2+1$ dimensional Abelian gauge theory with dynamics governed by the following $U(1)_M\times U(1)_{CS}$   Lagrangian
\be
L_{U(1) \times U(1) } = L_{MH} + L_{CSH} + L_{mix}
\label{uno}
\ee
Here
 $L_{MH}$  and $L_{CSH}$ are Maxwell-Higgs (MH)  and  Chern-Simons-Higgs (CSH) Lagrangians,
\be
L_{MH} = -\frac14 F^{\mu\nu}[A] F_{\mu \nu}[A] + \frac12|D_\mu[A]\phi|^2 - V_M[\phi]
\label{LMH}\ee
\be L_{CSH} = \frac\kappa4 \epsilon^{\alpha\beta\gamma} B_\alpha F_{\beta\gamma}[B] + \frac12|D_\mu[B]\eta|^2 - V_{CS}[\eta]
\label{LCSH}
\ee
with $(A_\mu,\phi)$ and $(B_\mu,\eta)$ the gauge and complex scalar fields in each sector. Field strengths and covariant derivatives for the MH sector are given by
\be
F_{\mu\nu}[A] = \partial_\mu A_\nu - \partial_\nu A_\mu \;, \;\;\;\; D_\mu\phi[A]=
(\partial_\mu +ie A_\mu)\phi
\ee
and an analogous formul\ae    for the $B,\eta$ fields in the CSH sector, with the  coupling constant $e$ replaced by $g$. Concerning the scalar potentials, we shall consider for the MH sector the usual quartic   symmetry breaking potential,
\be
V_M= \frac{\lambda_M}4 (|\phi|^2 -\phi_0^2)^2
\label{vm}
\ee
which for $\lambda_M = e^2/8$ (usually dubbed the Bogomolny point) exhibits, in the case of the ordinary Abelian Higgs model, first order self-dual equations \cite{Bogo}-\cite{deVS}.
%We shall also consider the case in which both the mass term and the $(|\phi|^2)^2$ term have the same sign for which symmetry is not spontaneously broken.

Concerning the  potential in the hidden sector, since the model is defined in $(2+1)$ space-time dimensions one has the choice
(on renormalizability grounds)  {to have a} fourth-order or sixth-order potential. We shall  {choose} the  {latter} option, since only in that case {do} first-order self-dual equations {exist} for the case of a Chern-Simons-Higgs theory with just one sector \cite{H}-\cite{JW}. We then have
\be
V_{CS} = \frac{\lambda_{CS}}4|\eta|^2(|\eta|^2 - \eta_0^2)^2
\label{vcs}
\ee
In the case  of an ordinary Chern-Simons-Higgs model the Bogomolny point corresponds to   $\lambda_{CS} = 4 g^4/\kappa^2$

The two sectors interact through  the mixing Lagrangian $L_{mix}${,} consisting of a BF-like coupling between gauge fields{,} and a Higgs portal coupling the two complex scalars{:}
\be
L_{mix} =
  \xi \varepsilon^{\mu \alpha \beta} A_\mu \partial_\alpha B_\beta +
 \zeta (|\phi|^2 -\phi_0^2)(|\eta|^2- \eta_0^2)
 \label{vmix}
\ee
We have chosen the Higgs portal in the way it arises in supersymmetric models with the same gauge dynamics in each sector (either MH-MH or CSH-CSH) \cite{AINS}-\cite{AIST}. As we shall see in the next section such kind of interaction also holds in the present case but of course supersymmetry forces the coupling constant $\zeta$ to take a particular value in terms of coupling constants $e,g,\xi$ and $\kappa$.

In our conventions{,}  gauge and scalar fields as well as gauge couplings $e,g$ have mass dimensions $1/2$   while the rest of the coupling constants have mass dimensions $1$.

The field equations associated to Lagrangian \eqref{uno} read in the Lorenz gauge
\be
  \Box A_\mu = i\frac{e}2(\phi^*\partial_\mu \phi - \phi \partial_\mu \phi^*) -e^2|\phi|^2 A_\mu - \xi\varepsilon_{\mu\alpha\beta}\partial^\alpha B^\beta
 \label{e1}
 \ee
 \be
  \kappa\varepsilon_{\mu\alpha\beta}\partial^\alpha B^\beta = i\frac{g}{2} (\eta^*\partial_\mu \eta - \eta \partial_\mu \eta^*)  -\,{g^2}\, |\eta|^2 B_\mu - \xi\varepsilon_{\mu\alpha\beta}\partial^\alpha A^\beta\label{e2}
  \ee
 \be
  \Box\phi + 2ie A_\mu\partial^\mu\phi - e^2 A_\mu A^\mu\phi = -\frac{\lambda_M}{2}(|\phi|^2- \phi_0^2)\phi +  \zeta(|\eta|^2- \eta_0^2)\phi \label{ef0}
  \ee
  \be
  \Box\eta + 2ig B_\mu\partial^\mu\eta - g^2 B_\mu B^\mu\eta = -\frac{\lambda_{CS}}{4}(|\eta|^4- \eta_0^4)\eta
    +  \zeta(|\phi|^2- \phi_0^2)\eta \label{ef}
\ee

Let us start by analyzing the Gauss laws resulting from eqs.\eqref{e1}-\eqref{e2},
\bea
&&\Box A_0 = j_0[\phi,A;e] -\xi\varepsilon_{ij}\partial^iB^j  \label{b1}\\
&&\kappa\varepsilon_{ij}\partial^iB^j = j_0[\eta,B;g] - \xi\varepsilon_{ij}\partial^iA^j\label{b2}
\eea
Here $i,j=1,2$ and the scalar currents are defined as
\be
j_\mu[\phi,A;e] = i\frac{e}2(\phi^*\partial_\mu \phi - \phi \partial_\mu \phi^*) -e^2|\phi|^2 A_\mu
\label{Gauss}
\ee
and an analogous formula for $j_\mu[\eta,B;g]$.

Integrating over space eqs.\eqref{b1}-\eqref{b2} we find
\bea
Q_A &=& \xi \Phi_B \label{f1}\\
Q_B &=& \xi\Phi_A +  \kappa\Phi_B \label{f2}
\eea
where we have defined electric charges and magnetic fluxes according to
\be
Q_A = \int d^2x j_0[\phi,A;e]\;, \;\;\;\; Q_B = \int d^2x j_0[\eta,B;g]\;,
\label{ana} \ee
\be \Phi_A = \int d^2x     \varepsilon_{ij}\partial^iA^j\;, \;\;\;\; \Phi_B = \int d^2x     \varepsilon_{ij}\partial^iB^j
\label{anal}
\ee
 It is interesting to note that finite energy electrically charged vortex configurations can only exist in the Abelian Higgs model when the gauge dynamics is governed by a Chern-Simons term. In the present case{,} vortices associated to the Maxwell term get a non trivial charge through the BF term that mixes both gauge fields without provoking any energy divergence, as we shall see below.

%%%%%%%%%%%%%%%%%%%%%%%%%%%%%%%%%%%%%%%%%%%%%%%%%%%%%%%%%%%%%%%%%%%%%%%%%%%%%%%%%%%%%%%%%%%%%%%%%%%%%%%
%%%%%%%%%%%%%%%%%%%%%%%%%%%%%%%%%%%%%%%%%%%%%%%%%%%%%%%%%%%%%%%%%%%%%%%%%%%%%%%%%%%%%%%%%%%%%%%%%%%%%%%
%%%%%%%%%%%%%%%%%%%%%%%%%%%%%%%%%%%%%%%%%%%%%%%%%%%%%%%%%%%%%%%%%%%%%%%%%%%%%%%%%%%%%%%%%%%%%%%%%%%%%%%
%%%%%%%%%%%%%%%%%%%%%%%%%%%%%%%%%%%%%%%%%%%%%%%%%%%%%%%%%%%%%%%%%%%%%%%%%%%%%%%%%%
%%%%%%%%%%%%%%%%%%%%%%%%%%%%%%%%%%%%%%%%%%%%%%%%%%%%%%%%%%%%%%%%%%%%%%%%%%%%%%%%%%%%%%%%%%%%%%%%%%%%%%%
%%%%%%%%%%%%%%%%%%%%%%%%%%%%%%%%%%%%%%%%%%%%%%%%%%%%%%%%%%%%%%%%%%%%%%%%%%%%%%%%%%%%%%%%%%%%%%%%%%%%%%%

Since we are interested in vortex solutions, we shall  make a static axially symmetric ansatz. As already {signalled}, in the case of the ordinary Maxwell-Higgs system one has to choose $A_0 =0$ since otherwise no finite energy solutions exist. In contrast, for the Chern-Simons-Higgs
 model  vortices should necessarily carry electric charged since otherwise the magnetic flux vanishes. In the present case eqs.~\eqref{f1}-\eqref{f2} force  to include both $A_0$ and $B_0$. Indeed, for static configurations $j_0$ as defined in eq.\eqref{anal} leads to an electric charge of the form
 \be
 Q_A =  -e^2\int d^2x |\phi^2|A_0
 \ee
so that taking $A_0 = 0$ would lead, in view of eq.\eqref{f1} to $\Phi_B=0$.
We then propose the following Ansatz
\be
\begin{array}{llll}
\phi= f(r) \exp(-in\varphi)\;,\;\; & A_\varphi = - \frac{1}{r}A(r)\;,\;\; & A_r = 0\;,\;\; &  A_0 = A_0(r) \\
\eta= q(r) \,\exp(-ik\varphi)\;,\;\; & B_\varphi = - \frac{1}{r}B(r) \;,\;\;& B_r= 0 \;,\;\; & B_0 = B_0(r)
\end{array}
\label{ans}
\ee
 With this Ansatz the   equations   for the spatial components of the gauge fields read
 \be
A''-\frac{A'}{r}+2r\xi B_0'-e(n+eA)f^2=0
 \label{e1r}
 \ee
 \be
\kappa B_0'+2\xi A_0'-\frac{g}{r}(k+g B)q^2=0
 \label{e2r}
 \ee
Concerning the Gauss laws, they take the form
\be
A_0''+\frac{A_0'}{r}+\frac{2\xi}{r}B'-e^2A_0 f^2=0
\ee
\be
B_0- \frac{2\xi A' +\kappa B'}{r g^2 q^2}=0
\ee
Finally, for the radial scalar field equations we have
\bea
&&f'' +\frac{f'}{r}-f\left(\frac{(n+eA)^2}{r^2}-e^2 A_0^2+\lambda_M\left(f^2-\phi_0^2\right)+2\zeta\left(\eta_0^2-q^2\right)\right)=0,
\nonumber\\
&&q''+\frac{q'}{r}-q\left(2\frac{(k+gB)^2}{r^2}-2g^2 B_0^2+4\zeta\left(\phi_0^2-f^2\right)+\lambda_{CS}\left(\left(\eta_0^2-2q^2\right)^2-q^4\right)\right)=0.\nonumber\\
 \label{radial}   \eea

 The appropriate boundary conditions for finite energy vortex solutions are
 \begin{equation}
\begin{array}{llllll}
A(0)= 0\;, &A_0(0) = \alpha_0\;, & \phi(0)=0\;,  & B(0)=0\;,& B_0(0) = \beta_0 & \eta(0)  = 0\\
A(\infty) = - {n}/e\;,& A_0(\infty) = 0\;,& \phi(\infty) = \phi_0\;, & B(\infty) = - {k}/g\;,&B_0(\infty) = 0 &\eta(\infty) = \eta_0\nonumber\\
\end{array}
 \end{equation}
 where $\alpha_0$ and $\beta_0$ are constants.

Concerning magnetic fluxes and electric charges we have from ansatz \eqref{ans}
\be
\Phi_A = ({2\pi}/e) n \;,  \;\;\;\;  Q_A = \xi  ({2\pi}/g) k
\label{con1}
\ee
\be
\Phi_B = ({2\pi}/g) k \;, \;\;\;\;  Q_B = \kappa  ({2\pi}/g) k   + \xi ({2\pi}/e) n
\label{con2}
\ee
We shall present the numerical solutions to the second order radial field equations \eqref{e1r}-\eqref{radial} in section 5. It should be stressed that no first order BPS equations can be found for the case of  Lagrangian $L_{U(1)\times U(1)}$ \eqref{uno}. Indeed, working \`a la Bogomolny it is not possible to accommodate the energy as a sum of squares plus a topological term. We shall then  proceed in the next section to  the  ${\cal N} = 2$ SUSY extension of a model with the same gauge dynamics searching for the existence of self-dual equations. As we shall see, this will imply the presence in the bosonic sector of an additional scalar field  with non-trivial dynamics, a scalar potential different from the one we introduced above and a reduction of the number of independent coupling constants. With all this{,} we shall see that Bogomolny first order equations can be found.
%%%%%%%%%%%%%%%%%%%%%%%%%%%%%%%%%%%%%%%%%%%%%%%%%%%%%%%%%
%%%%%%%%%%%%%%%%%%%%%%%%%%%%%%%%%%%%%%%%%%%%%%%%%%%%%%%%%
%%%%%%%%%%%%%%%%%%%%%%%%%%%%%%%%%%%%%%%%%%%%%%%%%%%%%%%%%
%%%%%%%%%%%%%%%%%%%%%%%%%%%%%%%%%%%%%%%%%%%%%%%%%%%%%%%%%
%%%%%%%%%%%%%%%%%%%%%%%%%%%%%%%%%%%%%%%%%%%%%%%%%%%%%%%%%
%%%%%%%%%%%%%%%%%%%%%%%%%%%%%%%%%%%%%%%%%%%%%%%%%%%%%%%%%

%%%%%%%%%%%%%%%%%%%%%%%%%%%%%%%%%%%%%%%%%%%%%%%%%%%%%%%%%%%%%%%%%%%%%%%%%%%%%%%%%%%%%%%%%%%%%%%%%%%%%%%
%%%%%%%%%%%%%%%%%%%%%%%%%%%%%%%%%%%%%%%%%%%%%%%%%%%%%%%%%%%%%%%%%%%%%%%%%%%%%%%%%%%%%%%%%%%%%%%%%%%%%%%
%%%%%%%%%%%%%%%%%%%%%%%%%%%%%%%%%%%%%%%%%%%%%%%%%%%%%%%%%%%%%%%%%%%%%%%%%%%%%%%%%%%%%%%%%%%%%%%%%%%%%%%
%%%%%%%%%%%%%%%%%%%%%%%%%%%%%%%%%%%%%     SECTION 3       %%%%%%%%%%%%%%%%%%%%%%%%%%%%%%%%%%%%%%%%%%%%%
%%%%%%%%%%%%%%%%%%%%%%%%%%%%%%%%%%%%%%%%%%%%%%%%%%%%%%%%%%%%%%%%%%%%%%%%%%%%%%%%%%%%%%%%%%%%%%%%%%%%%%%
%%%%%%%%%%%%%%%%%%%%%%%%%%%%%%%%%%%%%%%%%%%%%%%%%%%%%%%%%%%%%%%%%%%%%%%%%%%%%%%%%%%%%%%%%%%%%%%%%%%%%%%

\section{The ${\cal N} = 2$  supersymmetric   $U(1)\times U(1)$ Model}
We shall consider two ${\cal N} = 2$ supersymmetric sectors in $2+1$ dimensions, one with Maxwell-Higgs dynamics, the other one with Chern-Simons-Higgs dynamics with gauge symmetry breaking in both sectors. These two sectors will be coupled
using the CS-like superfield mixing introduced in \cite{AIST}.

Concerning the Maxwell-Higgs sector, we shall introduce the vector superfield $V_{M}$, which in the  the Wess-Zumino gauge is
composed   of a gauge field $A_\mu$, a 2-component complex spinor $\psi_h$, a real scalar field $M$  and an auxiliary scalar field $D$, and for the matter content, one introduces a scalar  superfield $\Phi=(\phi ,\lambda,F)$. containing a complex scalar $\phi$, a complex fermion $\lambda$ and an auxiliary field $F$ (which we  {will} omit as there is no superpotential).

The ${\cal N}=2$ supersymmetric Maxwell-Higgs action written in components reads  (see
\cite{AINS} and references {therein})
\begin{align}
S_{MH}  &=  \int d^3x \left(-\frac{1}{4}F_{\mu\nu}[A]F^{\mu\nu}[A] +
|D_\mu[A]\phi|^2+ \frac12 D^2+ e \left(|\phi|^2-\phi_{0}^2\right)D \right.\nonumber\\
&\left. + \frac12\partial_\mu M\partial^\mu M-e^2M^2|\phi|^2+ \frac{i}2 \bar{\psi}\slashed{D}[A]\psi
 + \frac{i}2 \bar\lambda \slashed{\partial} \lambda+ e M\bar{\psi}\psi
- {i \sqrt{2}e} \left(\bar{\psi}\lambda \phi-\phi^*\bar{\lambda}\psi\right)\right)\nonumber\\
\label{s1zzz}
\end{align}
We have included a Fayet-Iliopoulos term (last one in the first line) to implement gauge symmetry breaking. The $\gamma$-matrices are taken as the Pauli matrices with $\gamma^0 = \sigma^3,
\gamma^i = i \sigma^i$.

{The Action $S_{MH}$}  is invariant under the following
transformations with infinitesimal anticommuting  complex parameters $\eta$,
\begin{align}
\delta \phi= \sqrt{2} \bar{\eta}\psi\text{ , }&\delta\psi=-\sqrt{2}\eta \left(  i\gamma_\mu D^\mu[A] - e M \right) \phi\text{ , }
\nonumber\\
\delta M= i\left( \bar{\eta}\lambda-\bar{\lambda}\eta\right) \text{ , }&\delta A_\mu=i\left( \bar{\eta}\gamma_\mu \lambda-\bar{\lambda}\gamma_\mu \eta\right)   \text{ , }
\nonumber\\
\delta D=\partial^\mu\left(\eta\gamma_\mu\bar{\lambda}-\bar{\eta}\gamma_\mu\lambda \right) \text{ , }&\delta \lambda=\eta\left( i\epsilon^{\mu\nu\rho}\partial_\mu A_\nu \gamma_\rho+\slashed{\partial}M-i D\right)
\label{susy1zz}
\end{align}
%
%Note in particular that the Fayet-Iliopoulos term is indeed gauge invariant, its only contribution to the action is a linear term in $D$. It is also SUSY invariant, despite having no fermionic partner term, since the SUSY variation of $D$ is a total derivative.

In the case of the supersymmetric Chern-Simons-Higgs sector
  we also  introduce  a vector superfield $V_{CS}$,
composed (in the Wess-Zumino gauge)  of a gauge field $B_\mu$, a 2-component complex spinor $\chi$, a real scalar field $N$  and the auxiliary scalar field $d$.

Concerning  {the} matter content, one introduces a scalar  superfield $\Psi=(\eta,\sigma,f)$ containing a complex scalar $\eta$, a complex fermion $\sigma$ and an auxiliary field $f$ (again ignored).

The ${\cal N}=2$   supersymmetric action  written in terms of component fields   (see
\cite{AIST} and references there) takes the form
\begin{align}
  S_{CSH} &=\int d^3x \left(\kappa(\frac{1}{4}\epsilon^{\mu\nu\rho}B_\mu \partial_\nu B_\rho +\bar{\chi}\chi +  d N) +  |D_\mu[B] \eta|^2
+i \bar{\sigma} \gamma^\mu \not\!\!D_\mu[B] \sigma  -   g^2 N^2 \eta^2
\right.\nonumber\\
 & +\left. gd(|\eta|^2 - \eta_0^2)+  g N\bar{\sigma}\sigma  -i \sqrt{2}e (\bar\sigma \chi \phi + \phi^*
\bar \chi \sigma)  \right)
\label{lagrangianCSH}
\end{align}

Concerning   SUSY variations, one has
\begin{align}
\label{transfor2}
 \delta B_\mu=i\left( \bar{\epsilon}\gamma_\mu \chi -\bar{\chi}\gamma_\mu \epsilon \right) \text{ , } &
 \delta N=i\left( \bar{\epsilon}\chi -\bar{\chi}\epsilon \right)  \\
  \;\delta \eta = \sqrt{2}\bar{\epsilon}\sigma \text{ , }
 &  \delta d= \partial_\mu \left(\epsilon \gamma^\mu \bar{\chi} -\bar{\epsilon}\gamma^\mu \chi \right)  \\
\delta \chi=\epsilon\left( i\epsilon^{\mu\nu\rho}\gamma_\rho\partial_\mu B_\nu - id + \gamma^\mu\partial_\mu N \right)  \text{ , }
 &
{\delta \sigma=\sqrt{2}\epsilon(-i \gamma^\mu D_\mu[B] \eta+ g N\eta)}
%\label{transfor2}
\end{align}

Finally,  we choose, as the
 ${\cal N} = 2$ supersymmetric mixing between the two sectors, the one introduced in \cite{AIST} which reads:
 % superspace
%\begin{equation}
%S_{mix} = \chi \!\int \!d^3x\!\int\!d^2\theta d^2\bar\theta \, V_{CS}\bar D^\alpha D_\alpha V_{M}
%= \chi \!\int \!d^3x\!\int\!d^2\theta d^2\bar\theta \, V_{M}\bar D^\alpha D_\alpha V_{CS}
%\end{equation}
 in components
 \begin{equation}
S_{mix} = \xi\int d^3x \;\left(  \frac{1}{2}\epsilon^{\mu\nu\rho}B_\mu F_{\nu\rho}[A]+\left(\bar{\lambda}\chi+\bar{\chi}\lambda \right) +\left( DN+ dM\right)  \right)
\label{csportal}
\end{equation}
 The first term in \eqref{csportal} is a BF-like kinetic gauge mixing while the last two terms, once the auxiliary fields are eliminated using their equations of motion will give rise to a Higgs portal interaction.

The complete action governing the dynamics of our ${\cal N} = 2$ Maxwell-Chern-Simons-Higgs model is then defined as
\be
S_{{\cal N} = 2} =   S_{CSH} +   S_{MH}  + S_{int}
\label{7pepo}
\ee
with $S_{CSH}, S_{MH}, S_{int}$ given by eqs.\eqref{lagrangianCSH}, \eqref{s1zzz} and \eqref{csportal} respectively

To determine the explicit form of the symmetry breaking potential  in action \eqref{7pepo} we consider the equations associated to the auxiliary fields $D,d$ and to $N$
\begin{align}
\delta D:\text{ }&D=-e(|\phi|^2-\phi_0^2)-\xi N \nonumber\\
\delta d:\text{  }&\kappa N=-g(|\eta|^2-\eta_0^2) - \xi M\nonumber\\
\delta N:\text{  }&\kappa d =2g^2\eta^2 N - \xi D
\label{from}
\end{align}
Solving these algebraic equations we express  the values of fields   $D$ and $N$   in terms of the  remaining scalars, $\phi$, $\eta$ and $M$. Note that in the ordinary supersymmetric extension of the Maxwell-Higgs theory the $M$-field can be put to zero while in the present case it has non-trivial dynamics (see eq.\eqref{s1zzz})
Inserting the values of $D,d,N$ obtained in eq.\eqref{from} in action \eqref{7pepo} we get, for the symmetry breaking potential $V_{eff}(\phi,\eta,M)$
\begin{align}
V_{eff}(\phi,\eta,M)&=  \frac{g^2}{\kappa^2}\left(g(|\eta|^2-\eta_0^2) + \xi M\right)^2 |\eta|^2 \nonumber\\
& + \frac{1}{2}\left(e\left(|\phi|^2-\phi_0^2\right)-\frac{\xi}{\kappa}\left( g\left(|\eta|^2-\eta_0^2\right) + \xi M\right)\right) ^2 + e^2M^2|\phi|^2
\label{vef}
\end{align}
Compared with the potential chosen in the model of the previous section, composed of the {fourth} order symmetry breaking potential  \eqref{vm} for  the Maxwell-Higgs sector, the sixth order one \eqref{vcs}  for the Chern-Simons-Higgs sector and a Higgs portal term in  \eqref{vmix}, the resulting potential $V_{eff}$ exhibits two important differences. Primo, the coupling constants for each one of  the two terms in the potential are not independent parameters but, forced by SUSY, they are fixed in terms of the gauge coupling constants $e,g$ and the gauge mixing coupling constant $\xi$. Secundo, an additional real field $M$ enters in the bosonic sector Lagrangian   with couplings to the other  scalars.

In the case of the ordinary supersymmetric Maxwell-Higgs model the equations of motion of the real scalar $M$ has the solution  $M=0$ which is then chosen so   as to end with the Lagrangian of  the ordinary Abelian Higgs model in the bosonic sector. In contrast, the addition of  a Chern-Simons term  promotes its role to that of a field with non-trivial dynamics, as originally observed  in   \cite{M1}-\cite{M2}. The same phenomenon takes place in   the $U(1)\times U(1)$ model, in which Maxwell and Chern-Simons term belong to different sector which are connected through a BF gauge mixing term and a Higgs portal.

The potential $V_{eff}$ has a  minimum value $V_{eff}=0 $ for which  both $U(1)$ symmetries are broken
 \be
 V_{eff}\left[\vphantom{a^{1/2}} |\phi| = \phi_0, |\eta| = \eta_0, M=0\right] = 0
 \label{ssb}
 \ee
 Degenerate with this there is also  a second symmetric minimum  for $\xi \ne 0$
 \be
 V_{eff}\left[\vphantom{a^{1/2}}\eta = 0, \phi = 0, M = (g/\xi)\phi^2_0 \right] = 0
 \ee
provided   the following relation between $\phi_0$ and $\eta_0$ holds,
\be
e\phi_0^2 = -2g\frac\xi\kappa \eta_0^2
\ee
%%%%%%%%%%%%%%%%%%%%%%%%%%%%%%%%%%%%%%%%%%%%%%%%%%%%%%%%%%%%%%%%%%%%%%%%%%%%%

Concerning the Lagrangian for the bosonic sector, it takes the form
\bea
L_{bos}  &=&   -\frac{1}{4}F_{\mu\nu}[A]F^{\mu\nu}[A] +
|D_\mu[A]\phi|^2 +
 \frac12\partial_\mu M\partial^\mu M-e^2M^2|\phi|^2 +  \frac{\kappa}{4}\epsilon^{\mu\nu\rho}B_\mu \partial_\nu B_\rho  \nonumber\\
 && +  |D_\mu[B] \eta|^2
 + \frac12 \epsilon^{\mu\nu\rho}B_\mu F_{\nu\rho}[A] - V_{eff}[\phi,\eta,M]
\label{cac}
\eea
%%%%%%%%%%%%%%%%%%%%%%%%%%%%%%%%%%%%%%%%%%%%%%%%%%%%%%%%%%%%%%%%%%%%%%%%%%%%%
In order to obtain the BPS equations associated to this bosonic Lagrangian we can proceed as usual by equating to zero the SUSY variation for fermion fields \cite{WO}-\cite{ENS}. We get

\begin{align}
\epsilon^{ij}\partial_iA_j = \pm \left(
 -e(|\phi|^2 - \phi_0^2)  + \xi g (|\eta|^2 - \eta_0^2) + \xi M
\right)
\;, & \hspace{1 cm} D_\pm[A]\phi = 0 \nonumber
 \end{align}
 \begin{align}%\nonumber\\
  \epsilon^{ij}\partial_iB_j = \pm \frac1\kappa \left(
%\vphantom{\frac{\int^2}2}
\xi e\left( |\phi|^2 - \phi_0^2\right)  -\frac1\kappa\left( 2g^2\eta^2 + \xi^2\right) ) \left( g^2\left( |\eta|^2 - \eta_0^2\right)  + \xi M\right)
\right)
\;, & \hspace{0.4 cm} D_\pm[B] \eta = 0
\nonumber\end{align} %\nonumber\\
\begin{align}A_0=\pm M \;, & \hspace{1 cm} B_0 =\mp \frac{1}{\kappa} \left(g(|\eta|^2-\eta_0^2) + \xi M  \right)
\label{A0M}
\end{align}
with $D_\pm = D_1 \pm iD_2$. These equations, together with the Gauss law \eqref{Gauss} saturate the Bogomolny bound for the energy. The upper (lower)
sign corresponds to positive (negative) value of the magnetic charges.

%%%%%%%%%%%%%%%%%%%%%%%%%%%%%%%%%%%%%%%%%%%%%%%%%%%%%%%%%%%%%%%%%%%%%%%%%%%%%%%%%%%%%%%%%%%%%%%%%%%%%%%
%%%%%%%%%%%%%%%%%%%%%%%%%%%%%%%%%%%%%%%%%%%%%%%%%%%%%%%%%%%%%%%%%%%%%%%%%%%%%%%%%%%%%%%%%%%%%%%%%%%%%%%
%%%%%%%%%%%%%%%%%%%%%%%%%%%%%%%%%%%%%%%%%%%%%%%%%%%%%%%%%%%%%%%%%%%%%%%%%%%%%%%%%%%%%%%%%%%%%%%%%%%%%%%
%%%%%%%%%%%%%%%%%%%%%%%%%%%%%%%%%%%%%     SECTION 4       %%%%%%%%%%%%%%%%%%%%%%%%%%%%%%%%%%%%%%%%%%%%%
%%%%%%%%%%%%%%%%%%%%%%%%%%%%%%%%%%%%%%%%%%%%%%%%%%%%%%%%%%%%%%%%%%%%%%%%%%%%%%%%%%%%%%%%%%%%%%%%%%%%%%%
%%%%%%%%%%%%%%%%%%%%%%%%%%%%%%%%%%%%%%%%%%%%%%%%%%%%%%%%%%%%%%%%%%%%%%%%%%%%%%%%%%%%%%%%%%%%%%%%%%%%%%%

\section{Vortex solutions}
\begin{figure}[ptb]
\begin{subfigure}{.5\textwidth}
\centering
\includegraphics[width=0.9\linewidth]{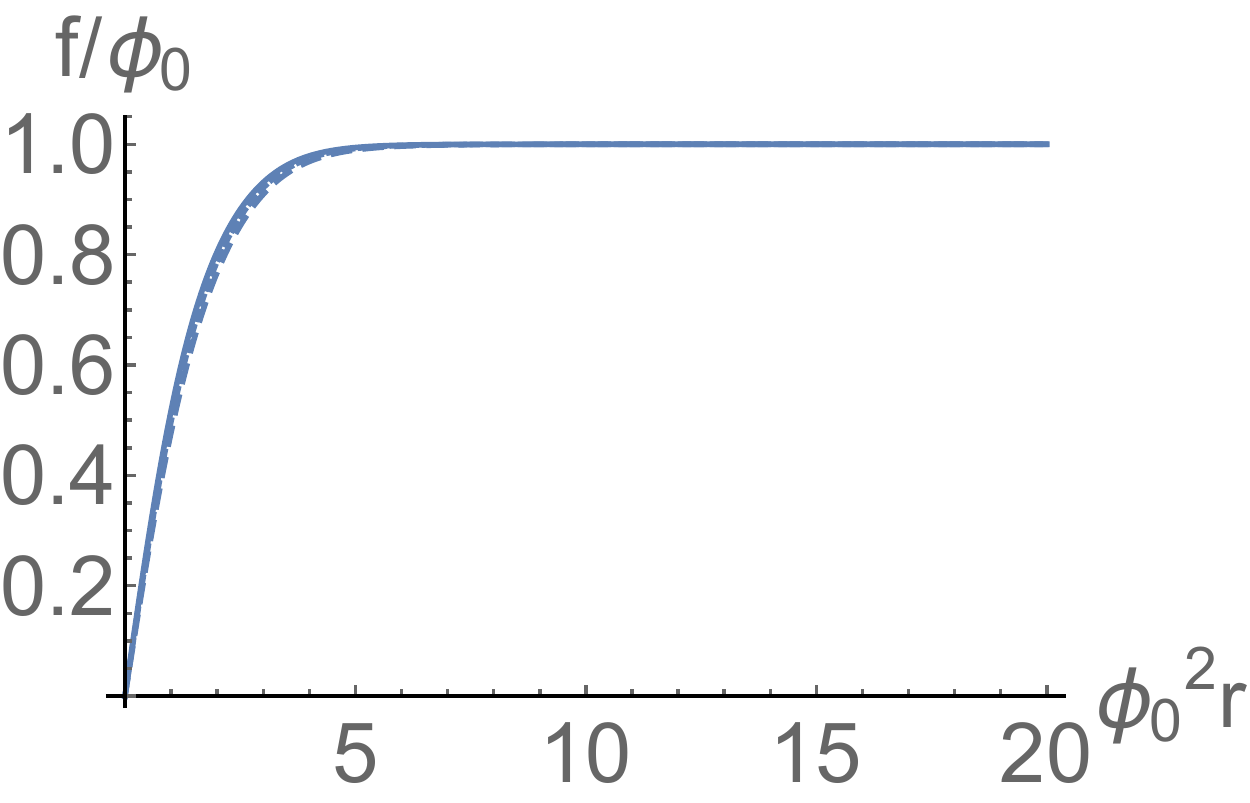}
\caption{}
\end{subfigure}
\begin{subfigure}{.5\textwidth}
\centering
\includegraphics[width=0.9\linewidth]{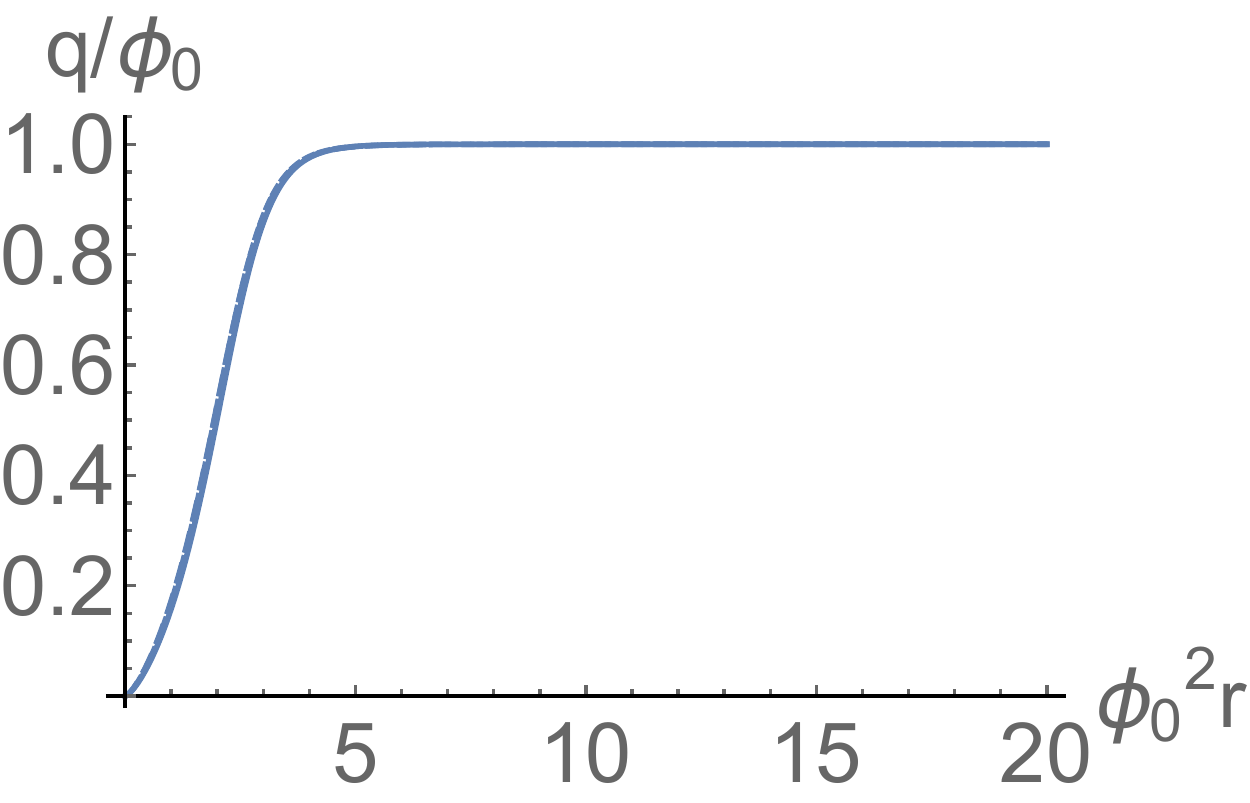}
\caption{}
\end{subfigure}
\begin{subfigure}{.5\textwidth}
\centering
\includegraphics[width=0.9\linewidth]{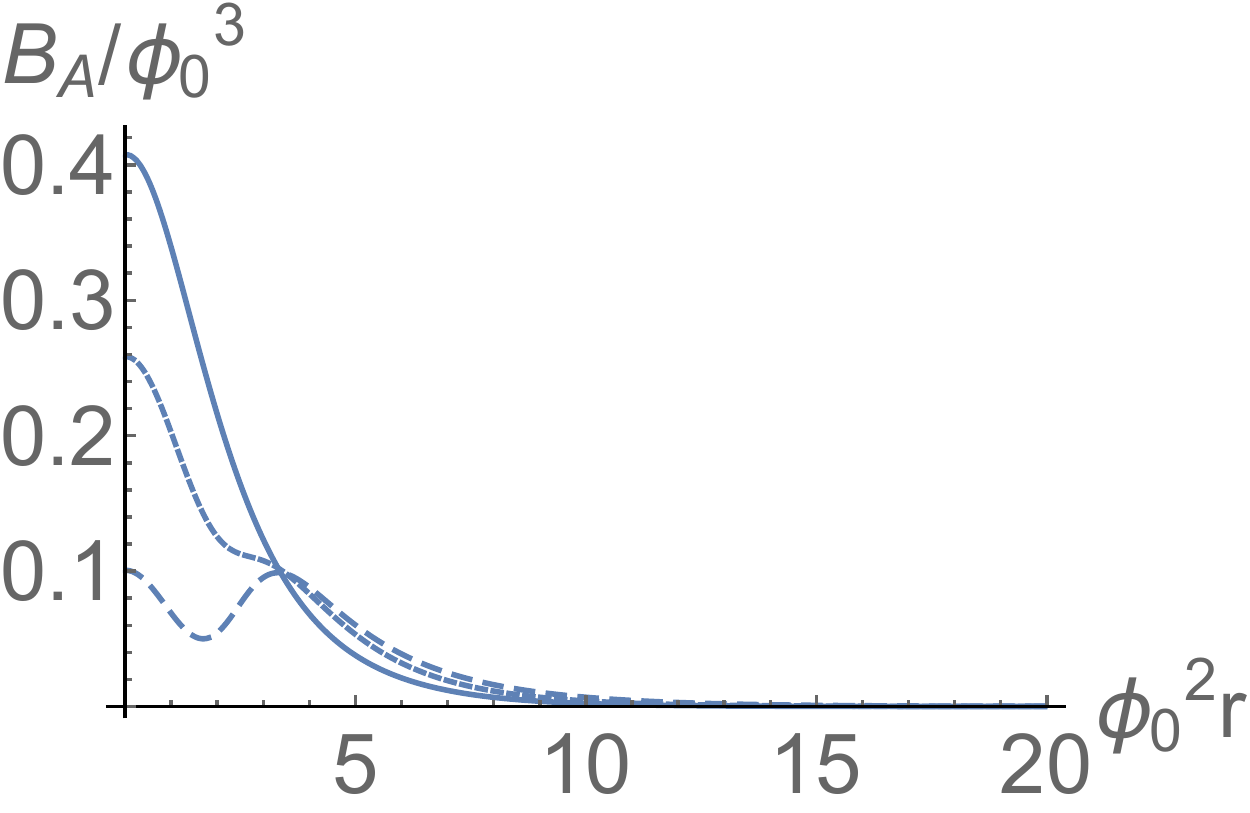}
\caption{}
\end{subfigure}
\begin{subfigure}{.5\textwidth}
\centering
\includegraphics[width=0.9\linewidth]{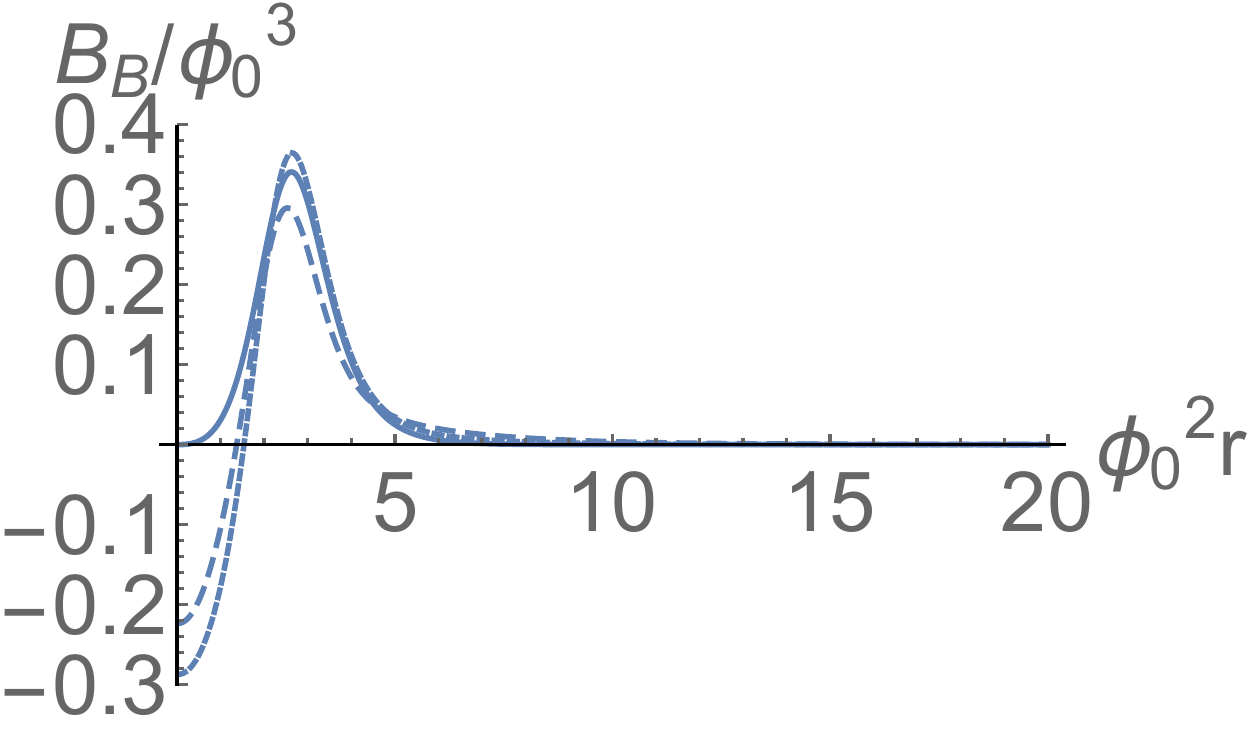}
\caption{}
\end{subfigure}
\begin{subfigure}{.5\textwidth}
\centering
\includegraphics[width=0.9\linewidth]{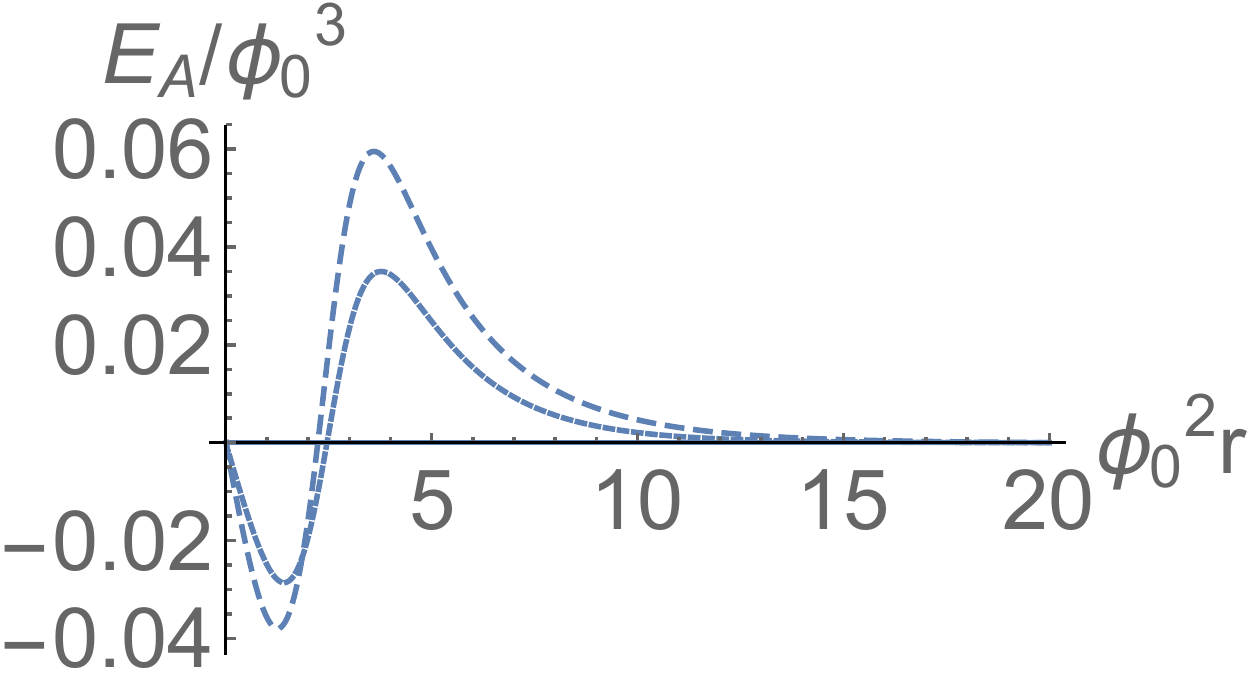}
\caption{}
\end{subfigure}
\begin{subfigure}{.5\textwidth}
\centering
\includegraphics[width=0.9\linewidth]{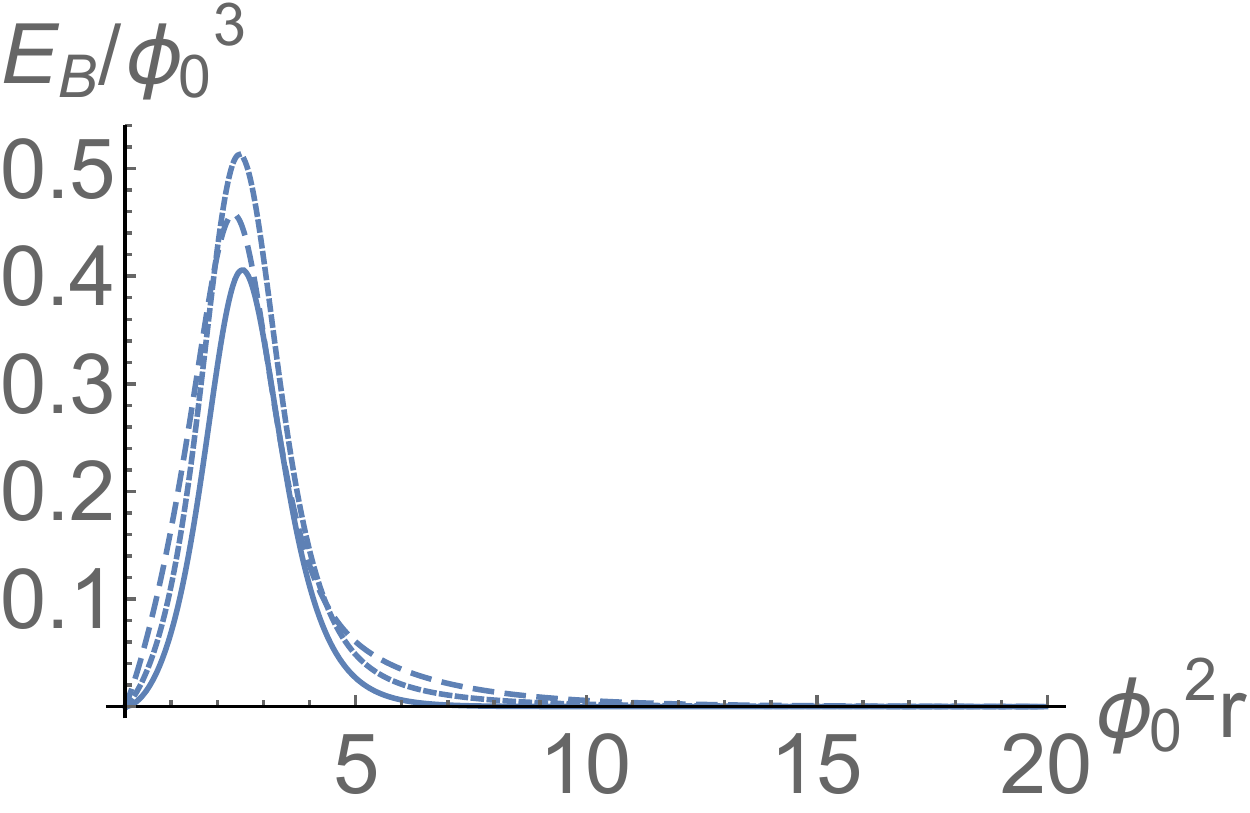}
\caption{}
\end{subfigure}
\caption{Field profiles for different values of the gauge mixing parameter, $\xi = 0.,0.1,0.2$.
Progressively larger dashing corresponds to larger values of $\xi$. Full line corresponds to the case of no mixing. All dimensional parameters are measured in units of $\phi_0$. We have chosen  $n=k= 1$ and $\eta_0 =\lambda_M = \lambda_{CS}=1$; $e= g=0.5$, $\zeta=0.1$, $\kappa=0.18$.    }
\label{fig1}
\end{figure}

\begin{figure}[ptb]
\begin{subfigure}{.5\textwidth}
\centering
\includegraphics[width=0.9\linewidth]{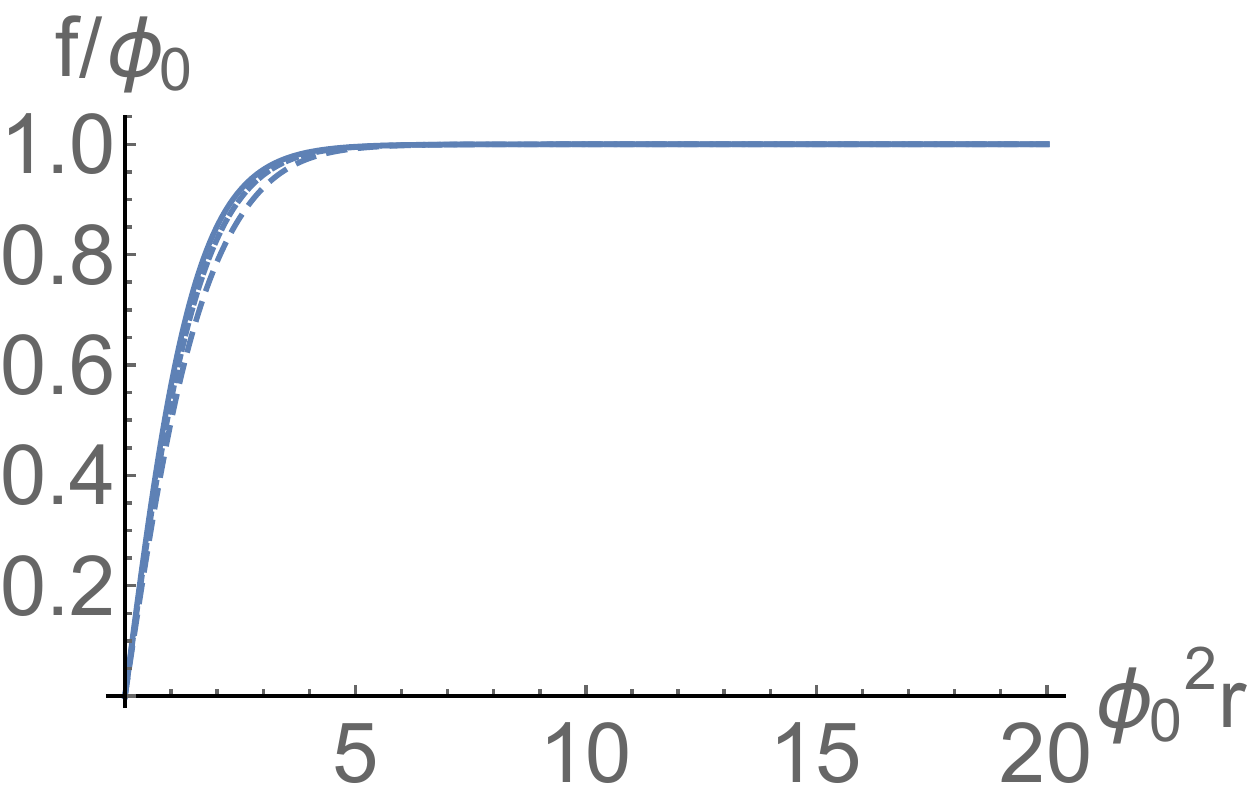}
\caption{}
\end{subfigure}
\begin{subfigure}{.5\textwidth}
\centering
\includegraphics[width=0.9\linewidth]{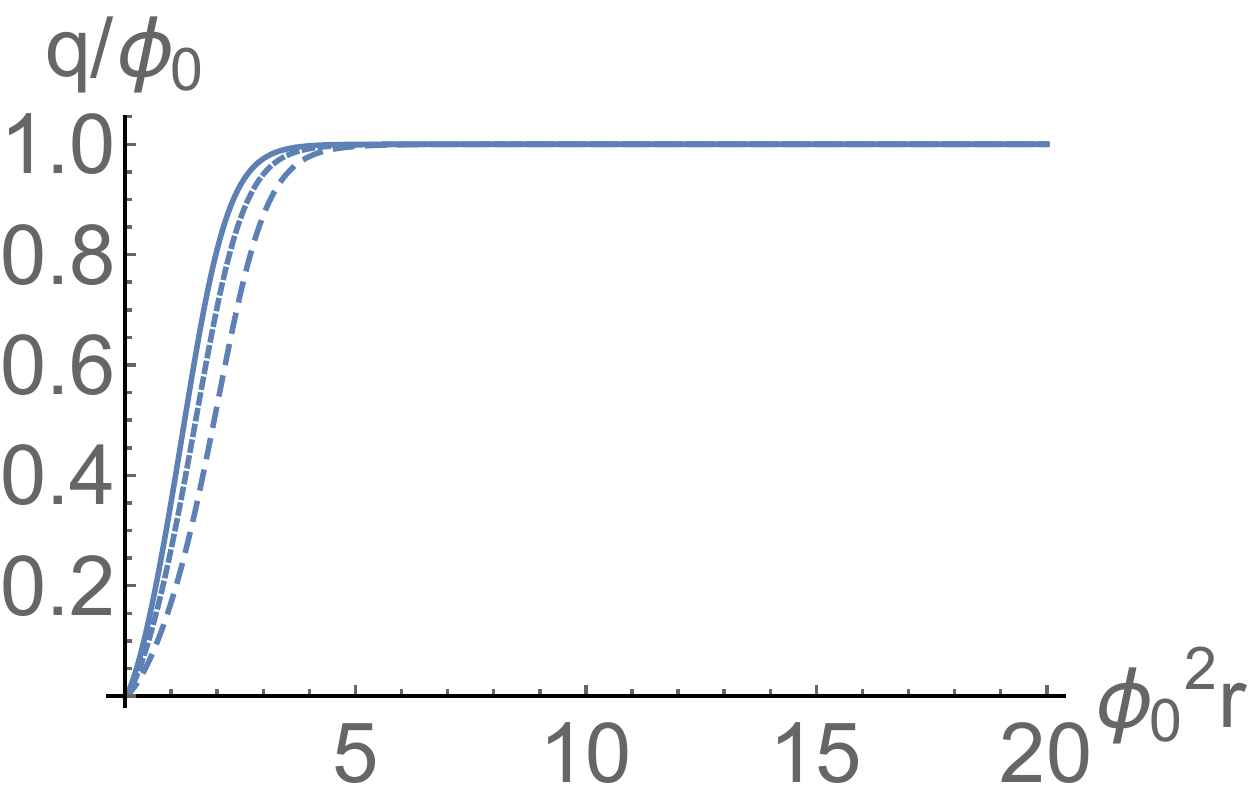}
\caption{}
\end{subfigure}
\begin{subfigure}{.5\textwidth}
\centering
\includegraphics[width=0.9\linewidth]{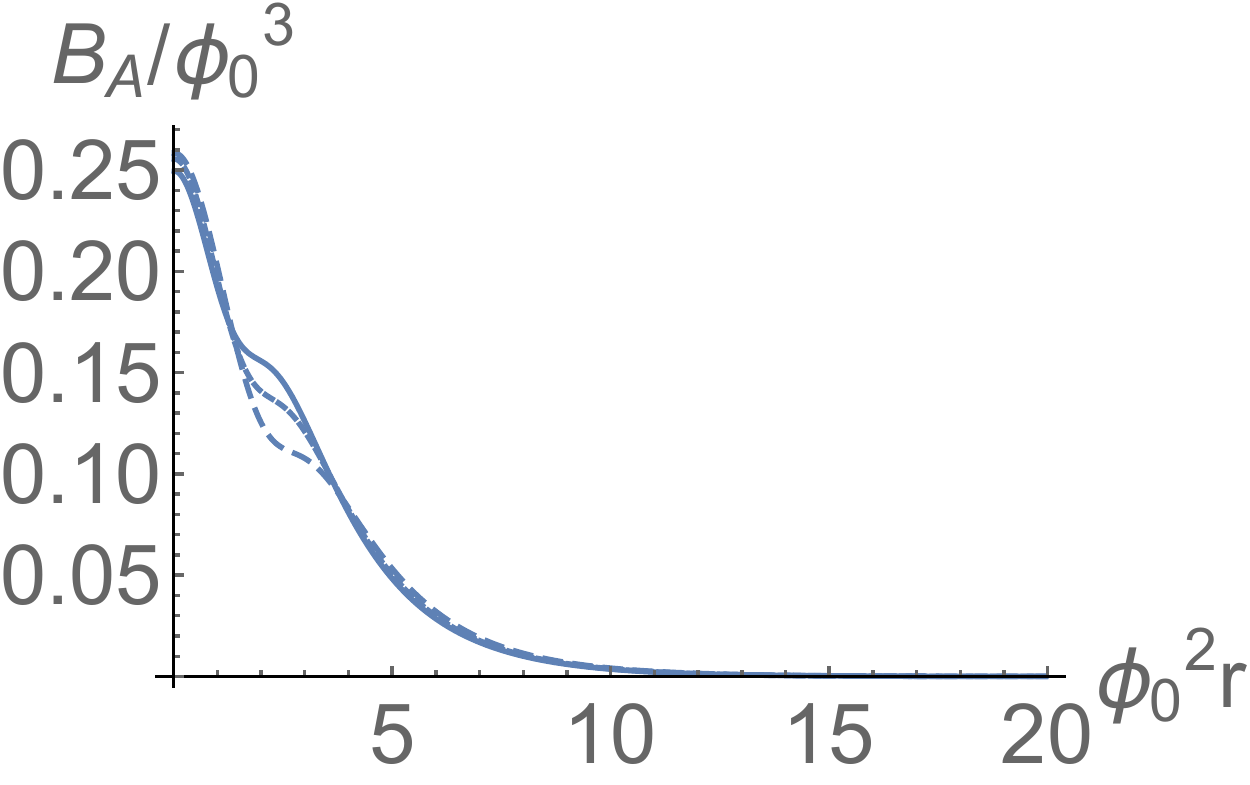}
\caption{}
\end{subfigure}
\begin{subfigure}{.5\textwidth}
\centering
\includegraphics[width=0.9\linewidth]{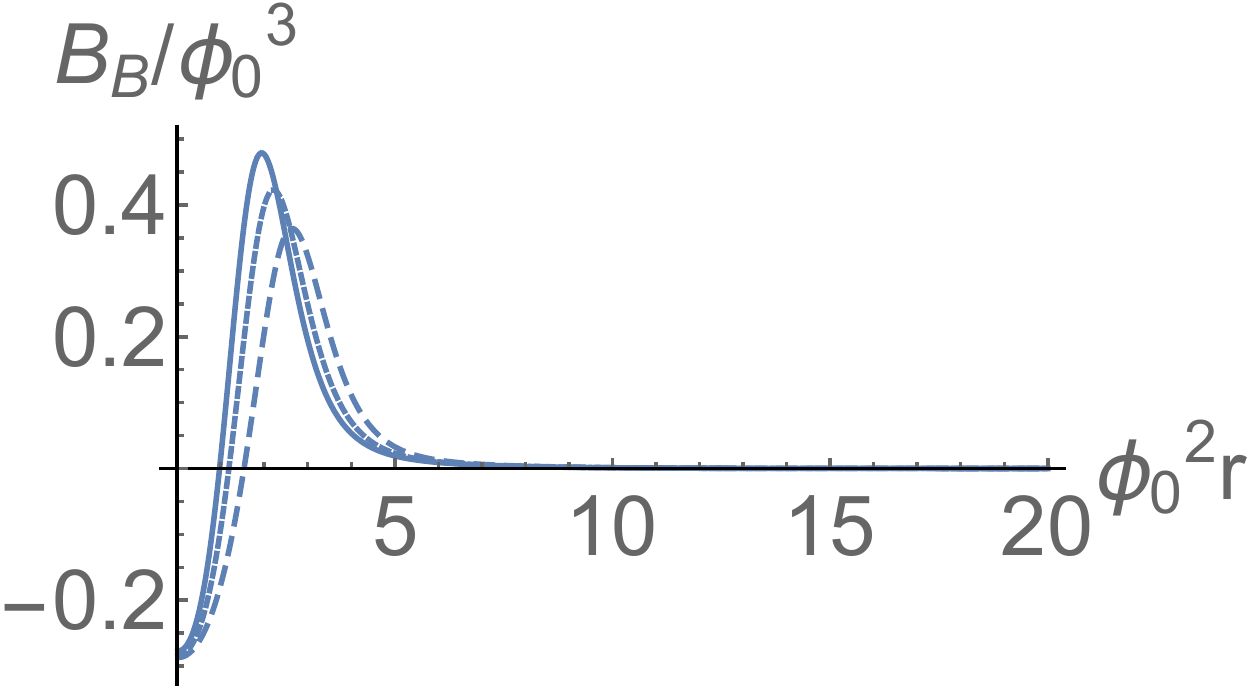}
\caption{}
\end{subfigure}
\begin{subfigure}{.5\textwidth}
\centering
\includegraphics[width=0.9\linewidth]{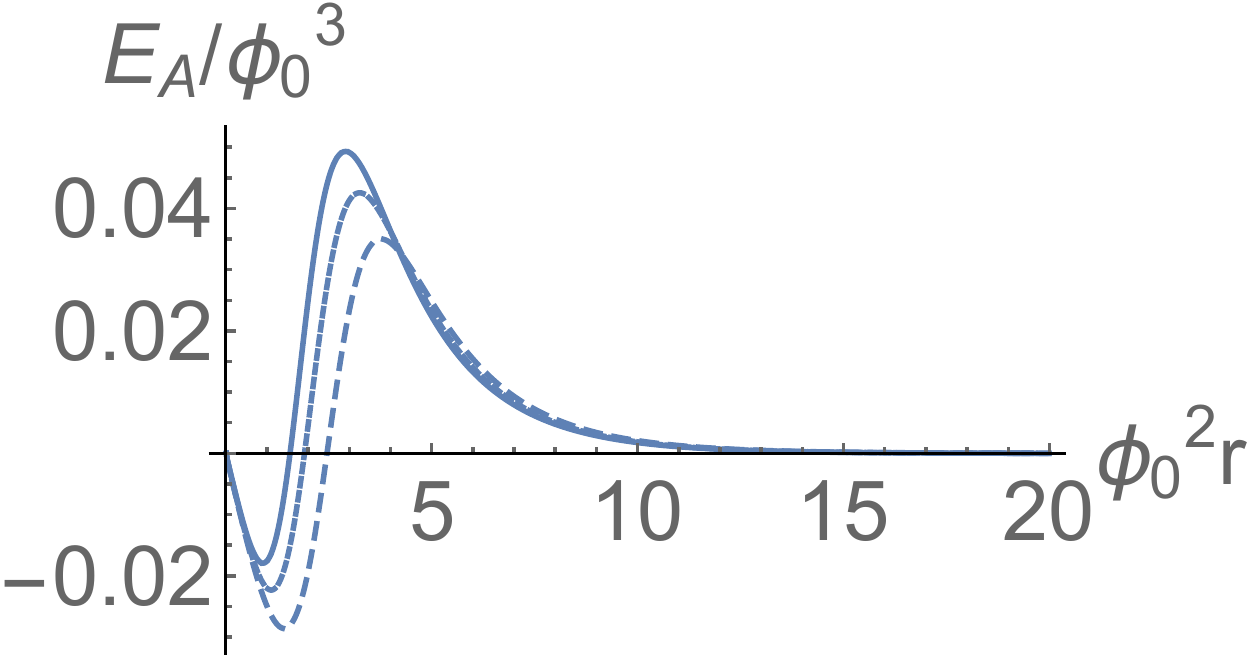}
\caption{}
\end{subfigure}
\begin{subfigure}{.5\textwidth}
\centering
\includegraphics[width=0.9\linewidth]{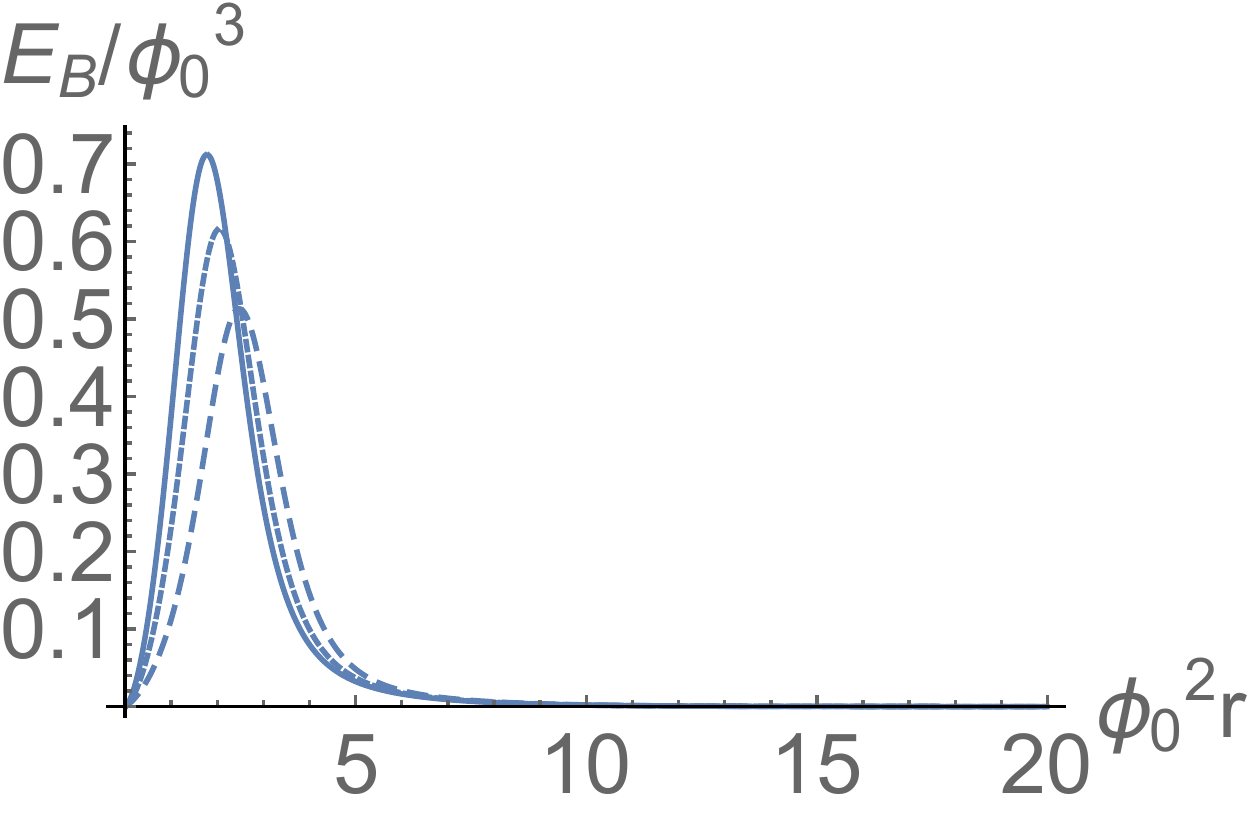}
\caption{}
\end{subfigure}
\caption{Field profiles for different values of the Higgs portal mixing, $\zeta = 0.$, $0.05$, $0.1$. Progressively larger dashing corresponds to larger values of $\zeta$. Full line corresponds to the case of no mixing. All dimensional parameters are measured in units of $\phi_0$. We have chosen  $n=k= 1$ and
 $\eta_0 =\lambda_M = \lambda_{CS}=1$, $e= g=0.5$, $\xi=0.1$, $\kappa=0.18$.}
\label{fig2}
\end{figure}

\begin{figure}[ptb]
\begin{subfigure}{.5\textwidth}
\centering
\includegraphics[width=0.9\linewidth]{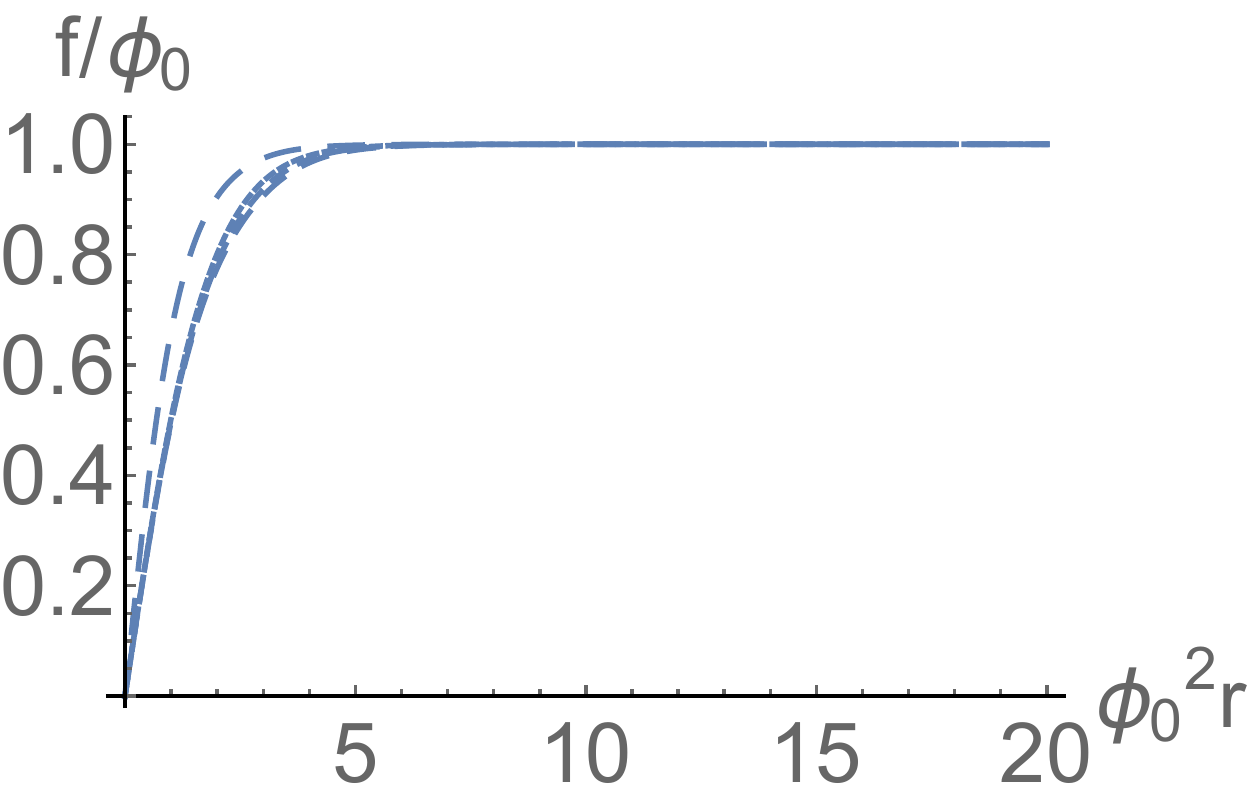}
\caption{}
\end{subfigure}
\begin{subfigure}{.5\textwidth}
\centering
\includegraphics[width=0.9\linewidth]{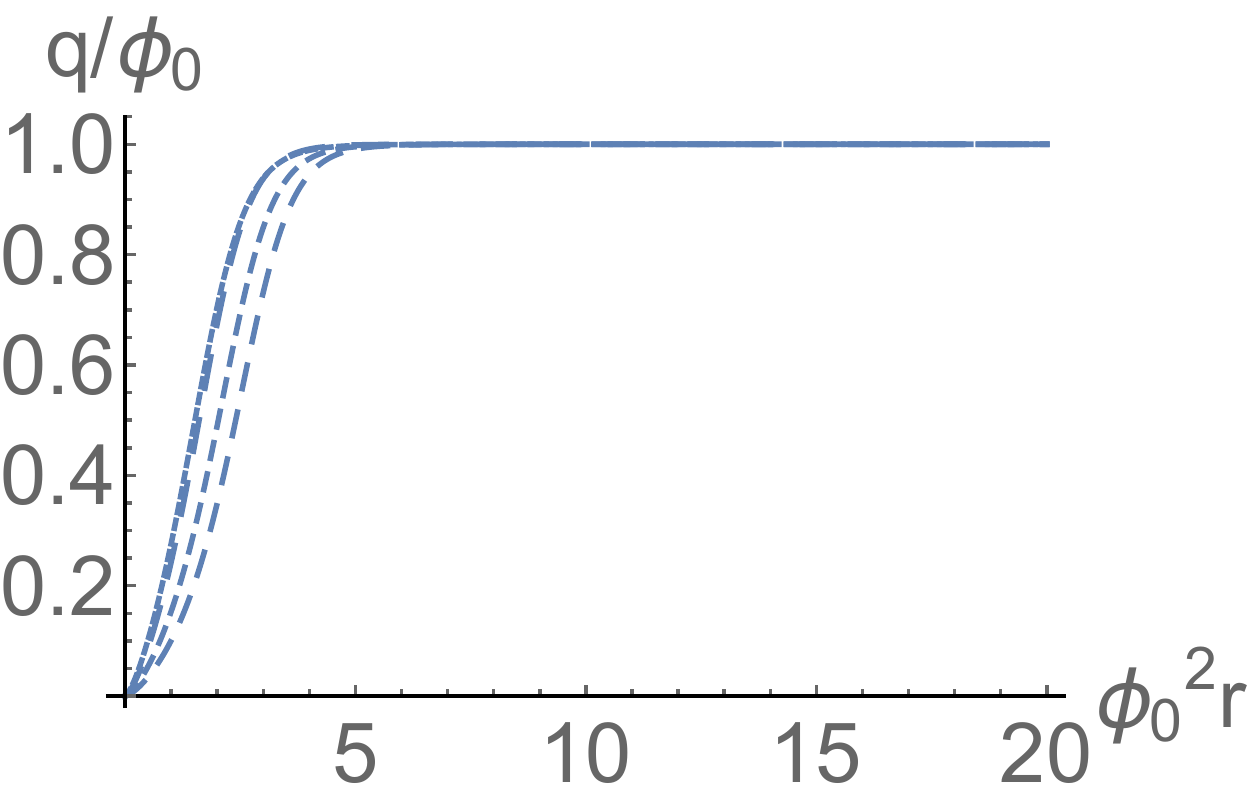}
\caption{}
\end{subfigure}
\begin{subfigure}{.5\textwidth}
\centering
\includegraphics[width=0.9\linewidth]{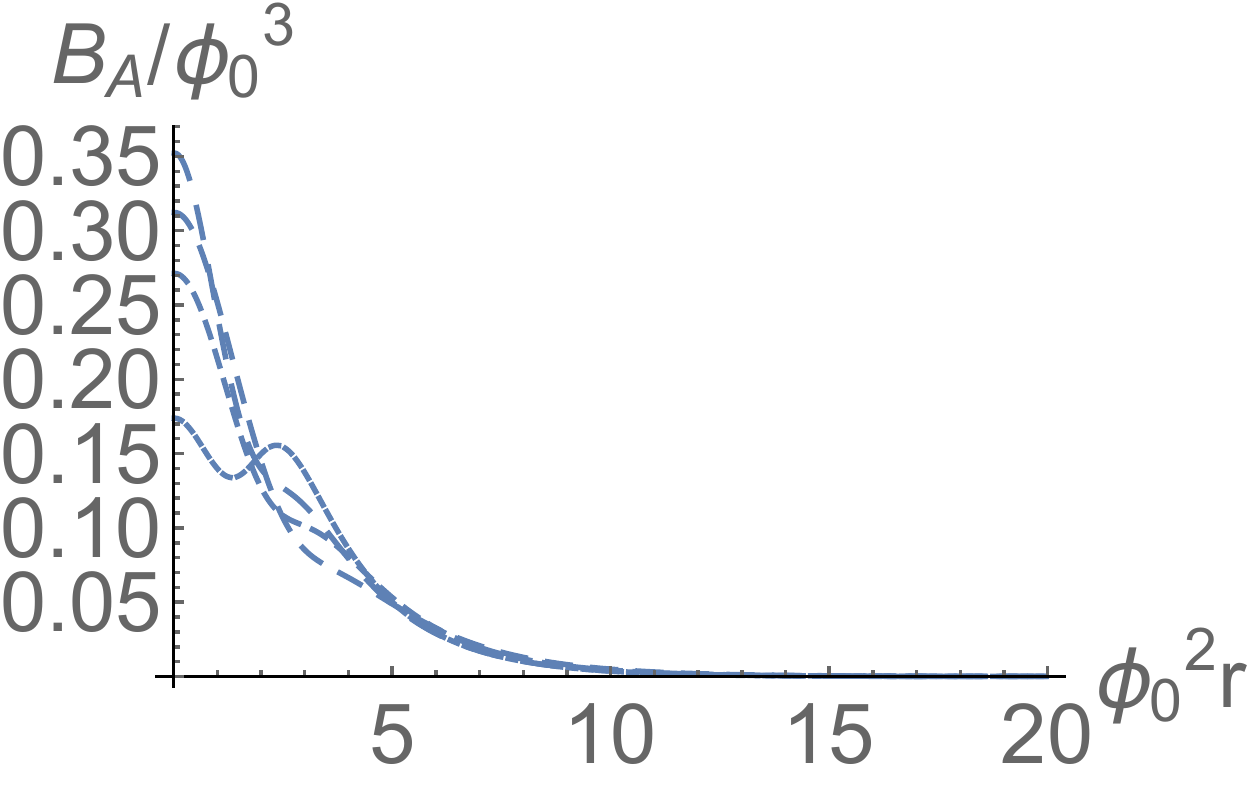}
\caption{}
\end{subfigure}
\begin{subfigure}{.5\textwidth}
\centering
\includegraphics[width=0.9\linewidth]{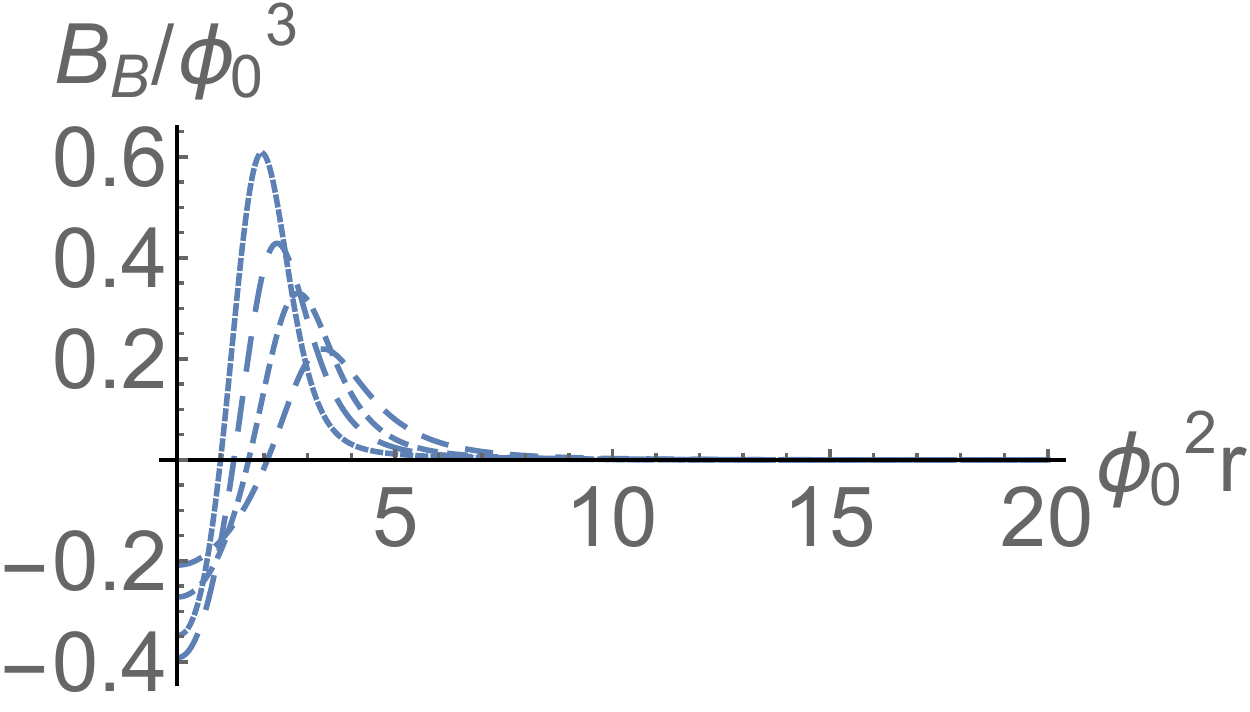}
\caption{}
\end{subfigure}
\begin{subfigure}{.5\textwidth}
\centering
\includegraphics[width=0.9\linewidth]{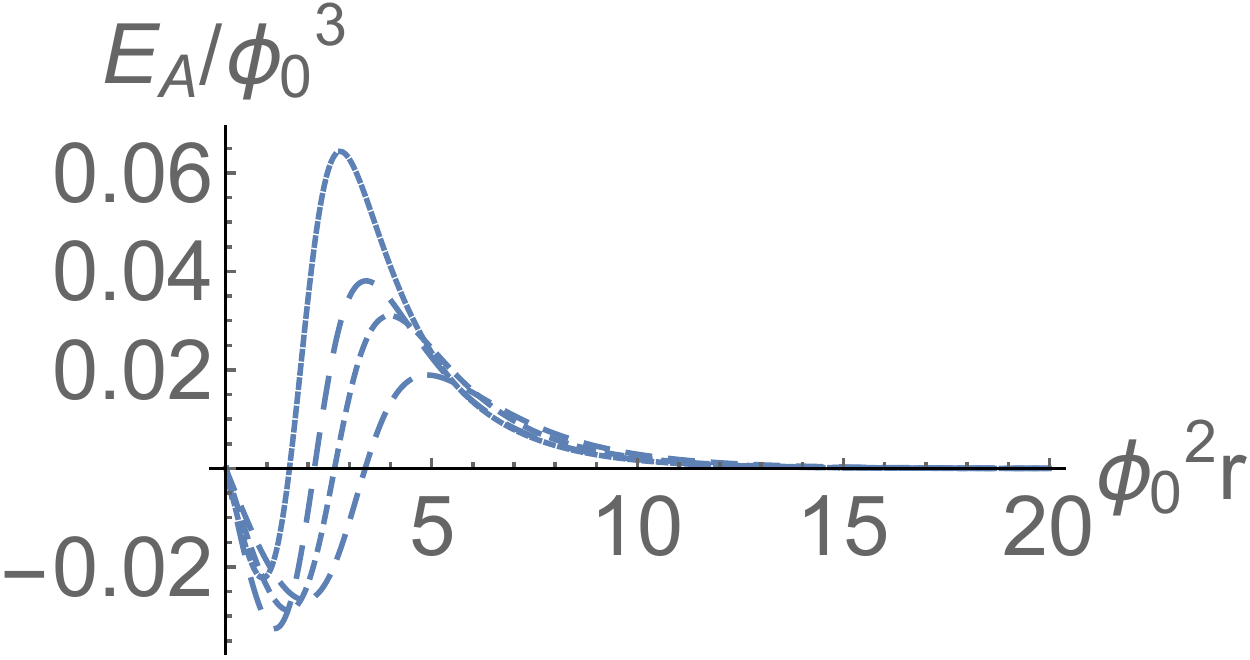}
\caption{}
\end{subfigure}
\begin{subfigure}{.5\textwidth}
\centering
\includegraphics[width=0.9\linewidth]{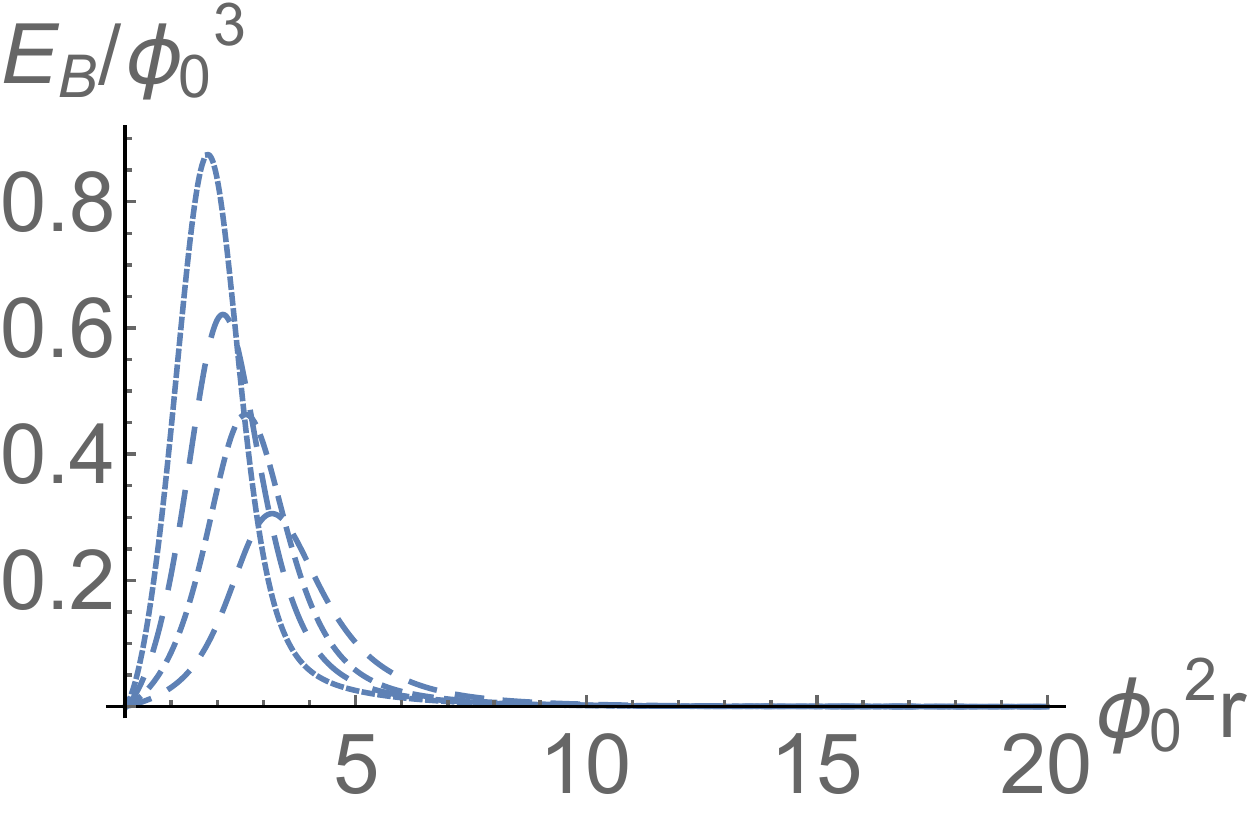}
\caption{}
\end{subfigure}
\caption{Field profiles for different values of the Chern-Simons coefficient, $\kappa = 0.1,0.2$, $0.3,0.4$.  Progressively larger dashing corresponds to higher $\kappa$. All dimensional parameters are measured in units of $\phi_0$. We have chosen  $n=k=1$, $\eta_0 =\lambda_M = \lambda_{CS}=1$ and $e= g=0.5$, $\xi = \zeta=0.1$}
\label{fig3}
\end{figure}

We first discuss the numerical solutions to the radial field  equations \eqref{e1}-\eqref{ef}  for the model with dynamics governed by Lagrangian \eqref{uno} introduced in Section 2. The numerical solver involves a second order central finite difference procedure with accuracy $\mathcal{O}(10^{-4})$. Field profiles of the obtained solutions for different ranges of parameters, are shown in figures \ref{fig1} to \ref{fig3}.

Figure 1 shows  the field profiles' dependence on the gauge mixing parameter $\xi$. One can see in figures 1(a) and 1(b) that the Higgs field profiles are almost insensitive to changes in $\xi$, growing from $0$ to its vacuum expectation values in the same way as in the ordinary MH and CSH models.
In contrast the $B_A$ magnetic field behavior drastically departs with respect to type-II superconductivity vortices. Indeed, one can see in figure 1(c) that as $\xi$ grows the shape of the profile changes and already for $\xi = 0.2$ (measured in units of $\phi_0$) it develops a second maximum which corresponds to a ring surrounding the origin.
This latter maximum is in fact larger than that in the core.

 Concerning   the electric field $E_A$, it is of course absent for $\xi = 0$ but, as $\xi$ grows,  it becomes non-trivial with the shape of  two rings  with opposite electric field signs . Also in the CS-Higgs sector the magnetic field  behavior radically changes: for $\xi=0$ $B_B$ vanishes at the origin  as in the ordinary CSH model. Now, for $\xi \ne 0$ it has a negative value near the origin and then it becomes positive as $r$ grows. In contrast the electric field $E_B$ has a different behavior as shown in figure 1(f): it vanishes at the origin independently of the $\xi$ value and solely the position and height of the maximum changes as $\xi$ varies. The  behavior of the electric fields can be better understood  by analyzing the bosonic sector of a supersymmetric version of the model, see next section.

Figure 2 shows the dependence of the field  on the value of the Higgs portal coupling constant $\zeta$ with the other parameters fixed. Again the Higgs field profiles do not exhibit a significant difference as $\zeta$ varies (figures 2(a) and 2(b)). Concerning the magnetic and electric fields their behavior is qualitatively similar to the previously discussed case{,} in which   the  $\xi$ value was changed.

In order to determine how the CS term affects the field behaviors we have analyzed their dependence on the Chern-Simons factor $\kappa$  keeping all other parameters fixed. As can be seen in figure 3 the behavior of the scalar field profiles shows a greater dependence on
$\kappa $ compared to that found when changing the mixing parameters. As for the magnetic and electric fields, the behavior is rather similar to those resulting from changing $\xi$ or $\zeta$, this showing that the presence of the CS term is one of the determinant factors in the physical properties of the model we are discussing.

We have also studied the dependence of the vortex solutions on coupling constants $e,g,\lambda_M, \lambda_{CS}$ without finding any notable change  when   their values were varied over a wide range.

  In the case of the ${\cal N} = 2$ SUSY model the search for axially symmetric vortex solutions for its bosonic sector should be done for boundary conditions  leading to the symmetry breaking vacuum (see eq. \eqref{ssb}) together with the condition $M(0) = A(0)$ and $M(r) \to 0$ as $r \to \infty$. We have solved the Bogomolny equations \eqref{A0M} associated to Lagrangian \eqref{cac} using  the same numerical approach as the one in the precedent section  finding that the profiles for the scalar, electric and magnetic fields do not differ significantly from the ones obtained for the model discussed in section 3. We then  solely present in figure 4 the results for the real scalar field $M$ whose presence was forced by supersymmetry.

\section{Summary and discussion}
In this work we have studied the case of  $U(1)\times U(1)$ gauge theories with  spontaneously broken symmetry in which the dynamics of one of the $U(1)$ gauge fields is dictated by a Maxwell term, the other one by a Chern-Simons term.
One can then consider that there two sectors, a visible   and a hidden one, as those in models discussed in the context of dark matter, supersymmetry breaking and other relevant problems in particle physics. Since the models are defined in $2+1$ space-time dimensions they can be taken  as toy models for realistic   models or  the high temperature limit of a realistic  quantum field theory.
 They could also be relevant in connection with condensed matter physics for the study of   superconductivity, quantum Hall effect  and topological insulators.

In section 2 we considered a purely bosonic theory in which the interaction between the two sectors include  a gauge field kinetic mixing  and a Higgs portal. An interesting point to notice has to do with the existence of electrically charged vortex solutions. As it is well known, the Abelian Higgs model with a Maxwell term does not support such solutions which in contrast should be there if the Chern-Simons action governs the gauge field dynamics. In the present case, because of the gauge mixing term, vortices in the Maxwell sector are necessarily charged while the charge in the CS sector gets a contribution from the flux in the other sector (eqs.\eqref{f1}-\eqref{f2}) this leading to the  relations \eqref{con1}-\eqref{con2} between quantized magnetic fluxes and electric charges in the two sectors.

The fact that we could not find self-dual equations associated to the $U(1)\times U(1)$ Lagrangian \eqref{uno} was  to be expected. Already for a $U(1)$ model including both  Maxwell and CS terms it was shown  in  \cite{M1} that an additional neutral scalar field should be added in order to be able  to write the vortex energy as a sum of squares plus a topological term which then gives the Bogomolny bound and the corresponding BPS equations.

Another way to find BPS equations is to proceed to the  ${\cal N}=2$ supersymmetric extension of a purely bosonic Lagrangian this implying the existence of the additional scalar field which in the case of the Maxwell-Chern-Simons dynamics    does  not have the trivial solution $M=0$  \cite{M2}. The same happens in the case of our $U(1)\times U(1)$ model  as was shown in section 3, where we found that $M$ couples not only with the complex scalar of its sector but also   with the hidden one leading to the effective potential
\eqref{vef}). Following the usual procedure we have obtained from the supersymmetric fermion field variations
    the BPS equations that, together with the Gauss law, allow to find the minima of the energy for each topological sector.

We present  in section 4 a detailed discussion of the numerical solutions that we have obtained for the two models considered in this work. One of the main features stemming from the gauge mixing of the two sectors is the onset of a purely radial electric field for the vortex configurations corresponding to the Maxwell sector because of the presence of a CS action in the other sector. In fact, the numerical results show   that the presence of the CS term is one of the determinant factors controlling the behavior of the system. As it is well-known, because of the parity anomaly in $2+1$ dimensions, Chern-Simmons terms  can be induced by
radiative quantum effects, even if they are not present as bare terms in the original Lagrangian. This is a key property in  $2+1$   bosonization \cite{FS}-\cite{FMS}  with many possible applications to condensed matter problems like those related to quantum Hall effect and topological insulators. In this context the effective action resulting from integration of fermions coupled to one of the gauge fields in $U(1) \times U(1)$ models could correspond to one of those studied in this work hence leading to topological configurations of the type described here.

Acknowledgments: G.T. is funded by Fondecyt grant No. 3140122 . F.A.S. is associated to CICBA and financially supported by PIP-CONICET, PICT-ANPCyT, UNLP and CICBA grants.

%%%%%%%%%%%%%%%%%%%%%%%%%%%%%%%%%%%%%%%%%%%%%%%%%%%%%%%%%%%%%%%%%%%%%%%%%%%%%%%%%%%%%%%%%%%%%%%%
%%%%%%%%%%%%%%%%%%%%%%%%%%%%%%%%%%%%%%%%%%%%%%%%%%%%%%%%%%%%%%%%%%%%%%%%%%%%%%%%%%%%%%%%%%%%%%%%
%%%%%%%%%%%%%%%%%%%%%%%%%%%%%%%%%%%%%%%%%%%%%%%%%%%%%%%%%%%%%%%%%%%%%%%%%%%%%%%%%%%%%%%%%%%%%%%%
%%%%%%%%%%%%%%%%%%%%%%%%%%%%%%%%%%%%%%%%%%%%%%%%%%%%%%%%%%%%%%%%%%%%%%%%%%%%%%%%%%%%%%%%%%%%%%%%%
%%%%%%%%%%%%%%%%%%%%%%%%%%%%%%%%%      FIGURES      %%%%%%%%%%%%%%%%%%%%%%%%%%%%%%%%%%%%%%%%%%%%%
%%%%%%%%%%%%%%%%%%%%%%%%%%%%%%%%%%%%%%%%%%%%%%%%%%%%%%%%%%%%%%%%%%%%%%%%%%%%%%%%%%%%%%%%%%%%%%%%%
%%%%%%%%%%%%%%%%%%%%%%%%%%%%%%%%%%%%%%%%%%%%%%%%%%%%%%%%%%%%%%%%%%%%%%%%%%%%%%%%%%%%%%%%%%%%%%%%%
%%%%%%%%%%%%%%%%%%%%%%%%%%%%%%%%%%%%%%%%%%%%%%%%%%%%%%%%%%%%%%%%%%%%%%%%%%%%%%%%%%%%%%%%%%%%%%%%%
%%%%%%%%%%%%%%%%%%%%%%%%%%%%%%%%%%%%%%%%%%%%%%%%%%%%%%%%%%%%%%%%%%%%%%%%%%%%%%%%%%%%%%%%%%%%%%%%%

%%%%%%%%%%%%%%%%%%%%%%%%%%%%%%%%%%%%%%%%%%%%%%%%%%%%%%%%%%%%%%%%%%%%%%%%%%%%%%%%%%%%%%%%%%%%%%%%%%%%%%%%
%%%%%%%%%%%%%%%%%%%%%%%%%%%%%%%%%%%%%%%%%%%%%%%%%%%%%%%%%%%%%%%%%%%%%%%%%%%%%%%%%%%%%%%%%%%%%%%%%%%%%%%%
%%%%%%%%%%%%%%%%%%%%%%%%%%%%%%%%%%%%%%%%%%%%%%%%%%%%%%%%%%%%%%%%%%%%%%%%%%%%%%%%%%%%%%%%%%%%%%%%%%%%%%%%
%%%%%%%%%%%%%%%%%%%%%%%%%%%%%%%%%%%%%%%%%%%%%%%%%%%%%%%%%%%%%%%%%%%%%%%%%%%%%%%%%%%%%%%%%%%%%%%%%%%%%%%%
%%%%%%%%%%%%%%%%%%%%%%%%%%%%%%%%%%%%%%%%%%%%%%%%%%%%%%%%%%%%%%%%%%%%%%%%%%%%%%%%%%%%%%%%%%%%%%%%%%%%%%%%
%%%%%%%%%%%%%%%%%%%%%%%%%%%%%%%%%%%%%%%%%%%%%%%%%%%%%%%%%%%%%%%%%%%%%%%%%%%%%%%%%%%%%%%%%%%%%%%%%%%%%%%%
%%%%%%%%%%%%%%%%%%%%%%%%%%%%%%%%%%%%%%%%%%%%%%%%%%%%%%%%%%%%%%%%%%%%%%%%%%%%%%%%%%%%%%%%%%%%%%%%%%%%%%%%
%%%%%%%%%%%%%%%%%%%%%%%%%%%%%%%%%%%%%%%%%%%%%%%%%%%%%%%%%%%%%%%%%%%%%%%%%%%%%%%%%%%%%%%%%%%%%%%%%%%%%%%%

\begin{figure}[ptb]

\centering
\includegraphics[width=0.5\linewidth]{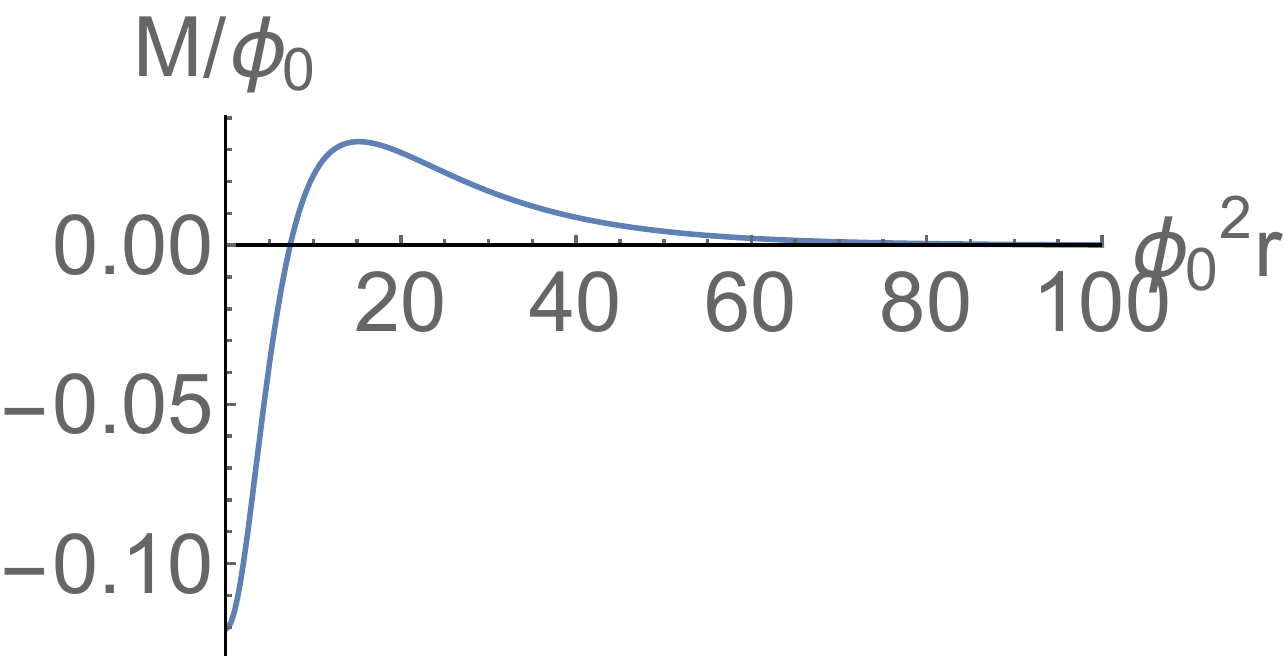}

\caption{Solution profile for the $M$ field. The solution corresponds to for  $n=k=1$, $\eta_0 =\lambda_M = \lambda_{CS}=1$ and $e= g=0.1$, $\kappa = 0.05$, $\xi=0.1$ (measured in units of $\phi_0$).}
\label{fig4}
\end{figure}

\end{document}